\documentclass[twocolumn,showpacs,aps,superscriptaddress]{revtex4-1}
\usepackage{amsmath}
\usepackage{amssymb}
\usepackage{bm}
\usepackage{graphicx}
\usepackage{dcolumn}
\usepackage{float}
\usepackage{natbib}
\usepackage[colorinlistoftodos,prependcaption]{todonotes}
\usepackage[pdftex,colorlinks=true,hypertexnames=false]{hyperref}

\begin{document}
\title{Order out of a Coulomb phase and Higgs transtion: frustrated transverse interactions of Nd$_2$Zr$_2$O$_7$}

\author{J. Xu}
\altaffiliation{jianhui.xu@helmholtz-berlin.de}
\affiliation{\mbox{Helmholtz-Zentrum Berlin f\"{u}r Materialien und Energie GmbH, Hahn-Meitner-Platz 1, D-14109 Berlin, Germany}}
\affiliation{\mbox{Institut f\"{u}r Festk\"{o}rperphysik, Technische Universit\"{a}t Berlin, Hardenbergstra{\ss}e 36, D-10623 Berlin, Germany}}

\author{Owen Benton}
\altaffiliation{john.benton@riken.jp}
\affiliation{\mbox{RIKEN Center for Emergent Matter Science (CEMS), Wako, Saitama, 351-0198, Japan}}

\author{A. T. M. N. Islam}
\affiliation{\mbox{Helmholtz-Zentrum Berlin f\"{u}r Materialien und Energie GmbH, Hahn-Meitner Platz 1, D-14109 Berlin, Germany}}

\author{T. Guidi}
\affiliation{\mbox{ISIS facility, Rutherford Appleton Laboratory, Didcot, OX11 0QX, UK}}

\author{G. Ehlers}
\affiliation{\mbox{Oak Ridge National Laboratory, Oak Ridge, P.O. Box 2008, Tennessee 37831, USA}}

\author{B. Lake}
\altaffiliation{bella.lake@helmholtz-berlin.de}
\affiliation{\mbox{Helmholtz-Zentrum Berlin f\"{u}r Materialien und Energie GmbH, Hahn-Meitner Platz 1, D-14109 Berlin, Germany}}
\affiliation{\mbox{Institut f\"{u}r Festk\"{o}rperphysik, Technische Universit\"{a}t Berlin, Hardenbergstra{\ss}e 36, D-10623 Berlin, Germany}}

\date{\today}

\begin{abstract}
The pyrochlore material Nd$_2$Zr$_2$O$_7$ with an ``all-in-all-out'' (AIAO) magnetic order shows novel quantum moment fragmentation with gapped flat dynamical spin ice modes. The parameterized spin Hamiltonian with a dominant frustrated ferromagnetic transverse term reveals a proximity to a U(1) spin liquid. Here we study magnetic excitations of Nd$_2$Zr$_2$O$_7$ above the ordering temperature ($T_\text{N}$) using high-energy-resolution inelastic neutron scattering. We find strong spin ice correlations at zero energy with the disappearance of gapped magnon excitations of the AIAO order. It seems that the gap to the dynamical spin ice closes above $T_\text{N}$ and the system enters a quantum spin ice state competing with and suppressing the AIAO order. Classical Monte Carlo, molecular dynamics and quantum boson calculations support the existence of a Coulombic phase above $T_\text{N}$. Our findings relate the magnetic ordering of Nd$_2$Zr$_2$O$_7$ with the Higgs mechanism and provide explanations for several previously reported experimental features.
\end{abstract}

\maketitle

\newcommand\ndzro{\mbox{Nd$_2$Zr$_2$O$_7$}}

\newcommand\tn{$T_\text{N}$}

\newcommand\tjx{\tilde{J}^{\tilde{x}}}
\newcommand\tjy{\tilde{J}^{\tilde{y}}}
\newcommand\tjz{\tilde{J}^{\tilde{z}}}

\newcommand\avampx{$\langle|\tau^{\tilde x}|\rangle$}
\newcommand\avampy{$\langle|\tau^{\tilde y}|\rangle$}
\newcommand\avampz{$\langle|\tau^{\tilde z}|\rangle$}

Competing interactions and geometrical frustration support highly degenerate states which suppress conventional magnetic order and lead to novel emergent states \cite{Lacroix2011book}. Classical spin ice (CSI) is a prominent example which is realized in (Dy/Ho)$_2$Ti$_2$O$_7$ pyrochlores consisting of a network of corner-sharing tetrahedra \cite{Gardner2010rev,Fennell2009,Morris2009,Harris1997,Henley2005}. In the CSI, the single-ion Ising anisotropy due to the crystal electric field (CEF) interactions frustrates the ferromagnetic (FM) interactions between the spins. This creates the ``2-in-2-out'' (2I2O) local constraint (ice rule) on  the spin configuration leading to infinite degeneracy on the pyrochlore lattice \cite{Harris1997}. However, an antiferromagnetic (AFM) interaction is not frustrated resulting in a long-range ``all-in-all-out'' (AIAO) order \cite{Lacroix2011book}.

With the introduction of transverse spin couplings, CSI transforms to quantum spin ice (QSI) allowing quantum tunnelling between the degenerate ice-rule states which realizes a type of U(1) quantum spin liquid \cite{Hermele2004,Onoda2010,Onoda2011,Shannon2012,Savary2012,Lee2012,Benton2012,Kato2015,Huang2018}. Recently, there has been a tremendous effort aimed at finding materials supporting QSI \cite{Gingras2014}. Several materials have been examined in the search for QSI including, for example, Yb$_2$Ti$_2$O$_7$ \cite{Ross2011}, Pr$_2$(Zr/Hf)$_2$O$_7$ \cite{Kimura2013,Sibille2018}, and Tb$_2$Ti$_2$O$_7$ \cite{Gardner1999}, but the evidence so far is ambiguous, complicated by multi-phase competitions, structural defects and low-lying crystal field levels \cite{Jaubert2015,Yan2017,Martin2017,Wen2017,Hamid2007,Princep2015}.

As a QSI candidate, \ndzro\ is an ideal material for modelling, having well-isolated, Kramers, Ising anisotropic, dipolar-octupolar CEF ground doublets and a clean, well-ordered structure and it has been under intensive study
\cite{Huang2014,Hatnean2015,Lhotel2015,Xu2015,Petit2016,Benton2016,Xu2016,Opherden2017,Lhotel2018,Xu2018,Xu2019}. Although it has an AIAO order as the ground state below $T_\text{N}\sim0.4$~K \cite{Lhotel2015,Xu2015}, it shows remarkably persistent spin dynamics \cite{Xu2016}, quantum moment fragmentation, gapped dynamical spin ice \cite{Petit2016,Xu2019}, gapped kagome spin ice in (111) fields \cite{Lhotel2018} and quantum spin-1/2 chains in (110) fields \cite{Xu2018}. The parameterized anisotropic pseudospin-1/2 $XYZ$ Hamiltonian indicates that the un-frustrated AFM $\tjz\approx-0.046$~meV induces the AIAO order, though the FM transverse $\tjx\approx0.09$~meV is approximately twice as strong as $|\tjz|$ \cite{Xu2019,Lhotel2018,Benton2016,Petit2016,Huang2014}. This is a result of the frustration for the FM $\tjx$ term. According to linear spin wave theory, the gap to the flat spin ice modes closes at $\tjx\//|\tjz|=3$ \cite{Benton2016} where spin ice configurations and AIAO order have the same energy. Classically, this signals the formation of an extensive ground-state manifold with icelike character for the $\tilde x$ component of the spin and the mixing of these states by quantum fluctuations caused by $\tjy$ and $\tjz$ stabilizes a U(1) spin liquid with dynamic emergent gauge fields \cite{Benton2016}. It was pointed out also that if there is a gapless Coulomb phase above \tn, the ordering transition would be a candidate for a Higgs transition \cite{Benton2016}.

\begin{figure*}[!htb]
\centering
\includegraphics[width=\textwidth]{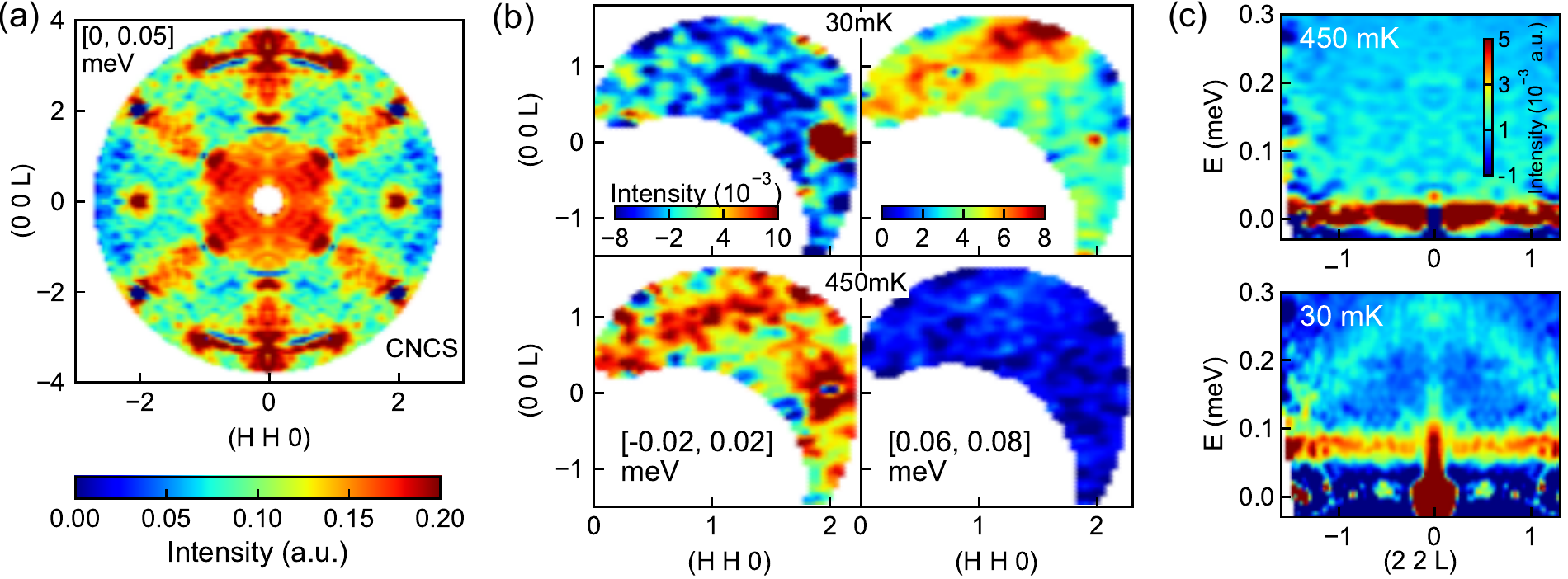}
\caption{(a) INS data measured on CNCS at 450\,mK integrated over [0, 0.05]\,meV. The data are averaged according to the symmetry of the reciprocal plane. (b) Constant-energy slices of the INS data measured on Osiris comparing the 30\,mK and 450\,mK data integrated over [-0.02, 0.02]~meV and [0.06, 0.08]~meV. (c) $E-Q$ slices along the (22L) direction in the reciprocal space of the Osiris data at 450~mK and 30~mK. The data at 20~K were subtracted for all the data. The strong sharp intensity at (220) at finite energy in the 30~mK Osiris data is an instrumental spurion resulted from leakage beyond the elastic channel.
\label{fig:ins_map}}
\end{figure*}

In this paper, we study the magnetic excitations of \ndzro\ above \tn\ using high-energy-resolution inelastic neutron scattering. We find that the gapped magnon excitations disappear and the pinch point pattern characteristic for spin ice correlations is still present but at zero energy. The gapless nature of the strong spin ice correlations points to a QSI state above \tn, which supports the theoretical speculations \cite{Benton2016}. Classical Monte Carlo simulations (MC) indicate that the system does not simply become paramagnetic with short-range AIAO correlations but enters a intriguing state with ice correlations for the $\tau^{\tilde x}$ spin components. Calculations for a finite temperature QSI with propagating monopole excitations, are qualitatively consistent with the scattering data. Our findings shed light on the anomalously slow spin dynamics probed by muon spin relaxation commonly seen in frustrated magnets \cite{Xu2016} and explains the puzzling temperature dependence of polarized neutron scattering data reported in Ref.~\cite{Petit2016}.

The \ndzro\ single crystal ($\sim2.5$~gram) was grown by using an optical floating zone furnace in the Core Laboratory for Quantum Materials (QM Core Lab) in Helmholtz-Zentrum Berlin (HZB) and characterized using X-ray powder diffraction and Laue diffraction \cite{Xu2019}. Inelastic neutron scattering experiments were conducted on the direct-geometry time-of-flight (tof) spectrometer CNCS at SNS in Oak Ridge National Lab and on the indirect-geometry tof spectrometer Osiris at ISIS Neutron Source in Rutherford Appleton Lab \cite{Ehlers2016,Telling2016}. For the CNCS measurement, the sample was mounted on a $^3$He insert which cooled the sample down to 240~mK. Neutrons of incident wavelength 4.98~\AA~(3.315\,meV) were used in the high-flux mode of the instrument (energy resolution $\sim0.1$~meV). Data were collected at 240\,mK, 450~mK and 20~K with a 360-degree sample rotation at a step of one degree. The large reciprocal space coverage provides an overview of the spin dynamics. To resolve the spin dynamics better, we performed the experiment on Osiris with a higher energy resolution 25\,$\mu$eV using the PG(200) analyser which analyses scattered neutrons of energy 1.84\,meV. The crystal was mounted onto a dilution refrigerator and data were collected at 30\,mK, 450\,mK and 20\,K. The data at 20\,K were used as the background for the two experiments. The software packages Dave \cite{dave}, Mantid \cite{mantid} and Horace \cite{horace} were used for data processing. Some data were averaged based on the symmetry of the scattering plane in order to improve the statistics. The classical Monte Carlo simulations and semiclassical molecular dynamics calculations were done using the MATLAB package SPINW \cite{spinw}, implementing our own codes (see supplementary materials \cite{SM}).

Figure~\ref{fig:ins_map}(a) shows the constant energy slice at 0.025~meV of the CNCS data at 450~mK (the 20~K data is subtracted as background). It is a well-formed strong pinch point pattern but at a much lower energy than in the ordered state ($\sim0.075$~meV) (see supplementary materials \cite{SM}). Fig.~\ref{fig:ins_map}(b) presents the background-subtracted data measured on Osiris at 30~mK and 450~mK with a much higher energy resolution 25~$\mu$eV. Integrating over [-0.02, 0.02]~meV, we see clearly a strong scattering arm along (111) of the pinch point pattern at 450~mK, consistent with the CNCS data, whereas the data at 30~mK does not show any signals except for the (220) magnetic Bragg peak. Conversely, in the data with integration over [0.06, 0.08]~meV, the 450~mK data does not show a clear pattern while the data at 30~mK shows the expected pinch-point spinwave modes from the AIAO order \cite{Petit2016,Xu2019}. In Fig.~\ref{fig:ins_map}(c), the $E-Q$ slices along the (22L) direction show that the gapped magnon excitations vanish above \tn\ and a strong scattering appears around the elastic line. In addition, a weak continuum at finite energy around (220) at 450~mK is also a new feature which has a similar intensity with the dispersing magnon modes at 30~mK.

\begin{figure}[!htb]
\centering
\includegraphics[width=\linewidth]{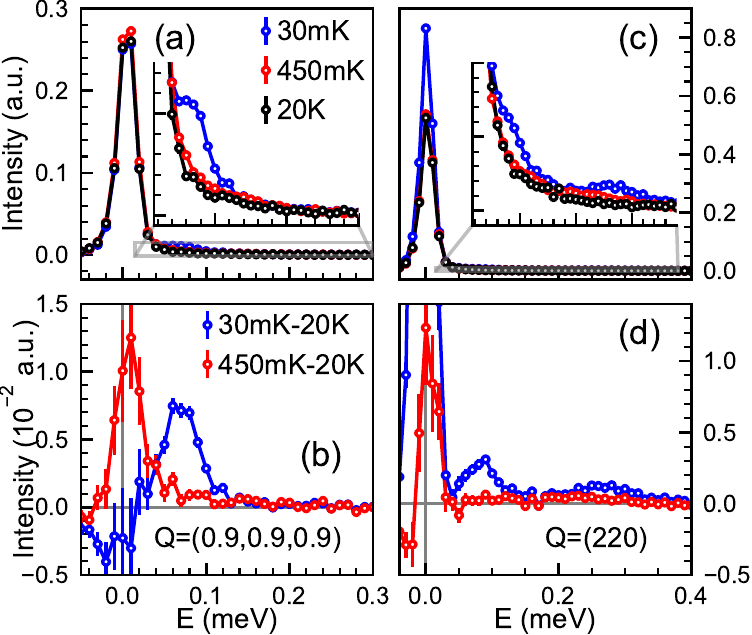}
\caption{One-dimentional intensity vs energy cuts through the Osiris data measured at 30\,mK, 450\,mK and 20\,K (upper panels) and with 30mK-20K and 450mk-20K subtractions (lower panels) at $Q=(0.9,0.9,0.9)$ [(a) and (b)] and $Q=(220)$ [(c) and (d)]. The grey lines in (a) and (b) indicate the regions plotted in the insets. The grey lines in (b) and (c) denote the zeros of the axes.}
\label{fig:ins_ie_cut}
\end{figure}

In Fig.~\ref{fig:ins_ie_cut}, we show the one-dimentional cuts through high-energy-resolution data at $Q=(0.9,0.9,0.9)$ and $(220)$ measured at 30~mK, 450~mK and 20~K. Comparing with the 20~K background data, we see pronounced gapped inelastic signals for 30~mK [besides the magnetic (220) Bragg peak] and strong elastic signals for 450~mK [Fig.~\ref{fig:ins_ie_cut}(a) and (c)]. After background subtraction as shown in Fig.~\ref{fig:ins_ie_cut}(b) and (d), we see that with the disappearance of the magnon excitations, most of the spectral weight changes to be elastic at 450~mK (there is a minor over-subtraction at negative energy transfers).

Therefore, we can conclude that above \tn, there appears a non-trivial paramagnetic phase with significant spin-ice correlations. It is quite surprising that the strong 2I2O correlations appears just above the AIAO ordering temperature though the broad feature around the AIAO Bragg peaks at [220] and [113] may indicate the existence of short-range AIAO order.

\begin{figure}[!htb]
\centering
\includegraphics[width=\linewidth]{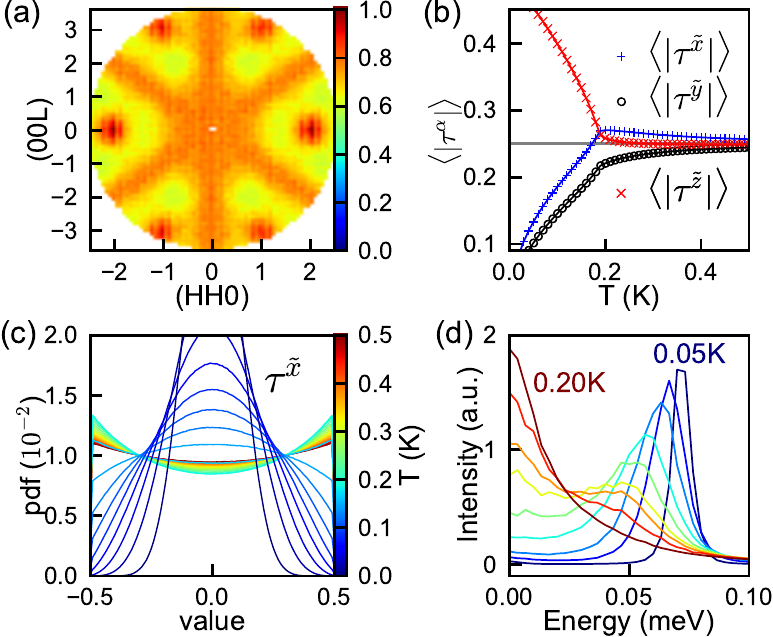}
\caption{Results of Monte Carlo simulations. (a) Neutron scattering structure factor (equal-time correlations) at 0.25\,K in the [HHL] reciprocal plane. (b) Temperature dependence of the thermal-average magnitudes of spin compositions (the grey horizontal line shows the expected result for a random spin configuration, $\tau/2$). (c) Probability distribution functions of $\tau^{\tilde x}$ (see Supplementary Materials \cite{SM} for corresponding figures for $\tau^{\tilde y}$ and  $\tau^{\tilde z}$). (d) Evolution of the gapped flat mode as a function of temperature (0.05, 0.1, 0.125, 0.15, 0.165, 0.175, 0.18, 0.185, 0.2~K) simulated using semi-classical molecular dynamics averaging over $Q$ from (0.1,0.1,0) to (0.9,0.9,0) \cite{SM}.}
\label{fig:mc_simu}
\end{figure}

To investigate the nature and origin of the high-temperature phase, we did classical Monte Carlo simulations based on the anisotropic spin Hamiltonian determined in Ref.~\cite{Xu2019},
\begin{equation}
{\cal H}_{XYZ}=
\sum_{\langle ij \rangle}
\big[
\tilde{J}_{\tilde{x}} \tilde{\tau}^{\tilde{x}}_i \tilde{\tau}^{\tilde{x}}_j
+
\tilde{J}_{\tilde{y}} \tilde{\tau}^{\tilde{y}}_i \tilde{\tau}^{\tilde{y}}_j
+
\tilde{J}_{\tilde{z}} \tilde{\tau}^{\tilde{z}}_i \tilde{\tau}^{\tilde{z}}_j
\big].
\label{eq:HXYZ}
\end{equation}
where $\tau^\alpha_i$ ($\alpha=x,y,z$) is the $\alpha$ component of the pseudospin-1/2 at site $i$ defined in the rotated local frames and ${\tilde J}_{\tilde \alpha}$ is the corresponding nearest-neighbor exchange constant \cite{Huang2014,Benton2016}. The calculations were done with a supercell of $6\times6\times6$ cubic unit cells, which has 3456 spins in total (for details see the supplementary materials \cite{SM}). The simulated specific heat indicates that the system enters the AIAO phase at $T_\text{N}^\prime\approx0.18$~K. Above $T_\text{N}^\prime$ at 0.25~K, the calculated neutron scattering structure factor [Fig.~\ref{fig:mc_simu}(a)] shows a pinch point pattern with broad signals around the AIAO Bragg peak wavevectors [220] and [113], which is quite consistent with the experiment.

To identify the spin correlations responsible for this novel scattering pattern, we calculated the thermal average of the amplitudes of the pseudospin components, $\langle|\tau^{\tilde\alpha}|\rangle$ ($\alpha=x,y,z$) and the probability distribution functions (pdfs) of $\tau^{\tilde\alpha}$ and the average over tetrahedra $1/4\sum_\boxtimes\tau^{\tilde\alpha}$. Fig.~\ref{fig:mc_simu}(b) shows the temperature dependence of $\langle|\tau^{\tilde\alpha}|\rangle$. At base temperature, only \avampz\ are significant consistent with the AIAO order. With increasing the temperature, \avampz\ decreases while \avampx\ and \avampy\ increase due to enhanced thermal fluctuations and all approach $\tau/2=0.25$ ($\tau=1/2$ is the amplitude of pseudospin) expected for completely random spin configuration. Remarkably, above $T_\text{N}^\prime$, \avampx\ gets higher than the thermal average $\tau/2$ and \avampz\ and then decays slowly with increasing temperature exhibiting a maximum around $T_\text{N}^\prime$. By contrast, \avampz\ decreases quickly and even gets slightly lower than $\tau/2$, revealing the breakdown of the AIAO correlations.

The correlations of $\tau^{\tilde x}$ is the origin of the pinch point pattern. We characterize the correlations of $\tau^{\tilde x}$ with the probability distribution functions mentioned above. As shown in Fig.~\ref{fig:mc_simu}(c), at temperatures below $T_\text{N}^\prime$, pdf$(\tau^{\tilde x})$ is a Gaussian function centered at zero and its width increases with raising temperature. Above $T_\text{N}^\prime$, it surprisingly turns to be two peaks at $\pm\tau$ decaying with further increasing temperature, which means that $\tau^{\tilde x}$ points into the tetrahedron for half of the spins and out of the tetrahedron for the other half. The pdf of the average on tetrahedra $1/4\sum_\boxtimes\tau^{\tilde x}$ shows how the half/half in/out $\tau^{\tilde x}$ is distributed on the tetrahedra which is always a Gaussian function centered at zero (shown in Ref.~\cite{SM}). The above two statistical quantities indicate that ice-rule correlations appear for the $\tau^{\tilde x}$ component on the tetrahedra which is consistent with the FM $\tjx$. On the other hand, pdf$(\tau^{\tilde z})$ and pdf$(1/4\sum_\boxtimes\tau^{\tilde z})$ change from peaks at either $\tau$ or $-\tau$ (depending on the AIAO domain type) to be a broad Gaussian peak centered at zero indicating the loss of the AIAO correlations of $\tau^{\tilde z}$ above $T_\text{N}^\prime$ \cite{SM}.

\begin{figure}[!tb]
\centering
\includegraphics[width=\linewidth]{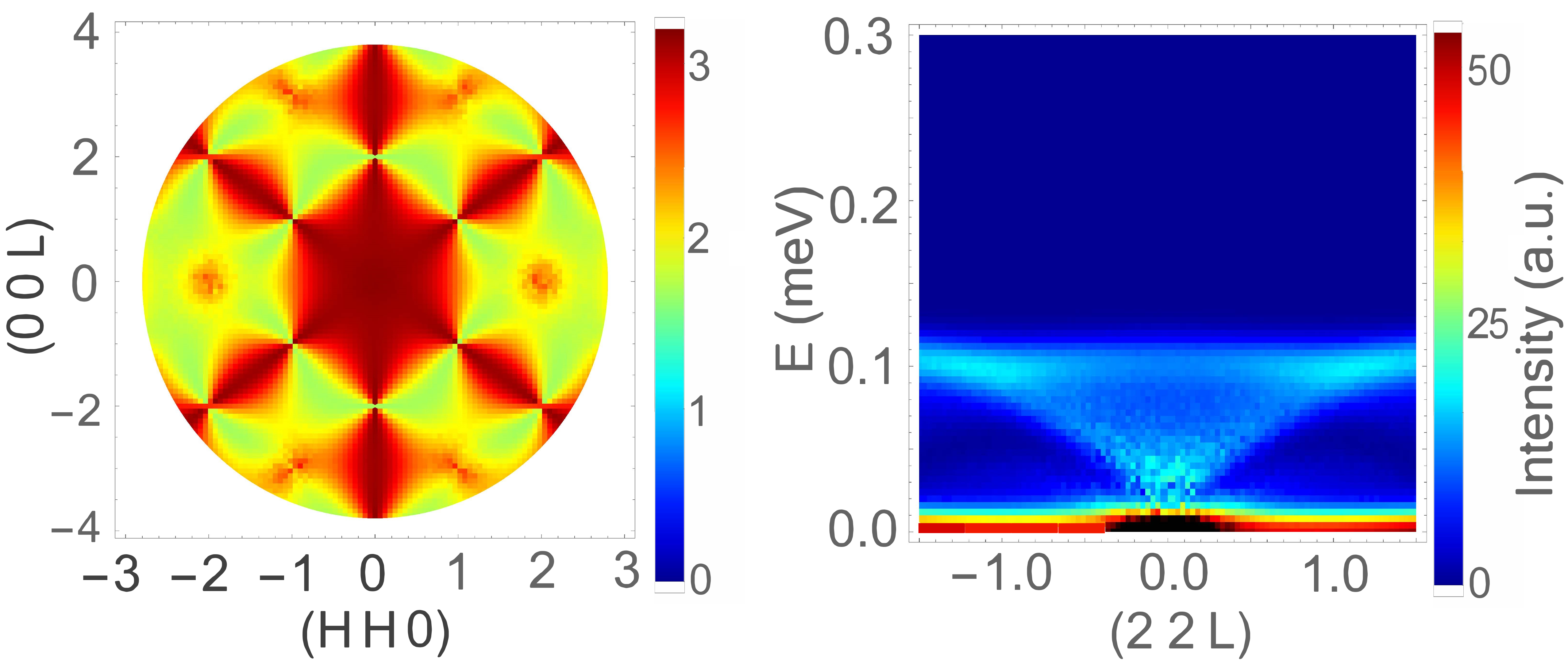}
\caption{(a) Calculated neutron scattering pattern at 450~mK (integrated over [0, 0.05]~meV) of the magnetic Coulomb phase and monopoles using the Hamiltonian of \ndzro\ where the pinch point pattern is caused by the divergence-free Coulomb phase and the broad signals at $Q=[220]$ and $[113]$ are due to monopoles. (b) $E-Q$ plot along (22L) showing the scattering continuum due to monopole hopping (elastic) and creation (inelastic) at 450~mK. Energy resolutions 0.1~meV and 0.02~meV were included for (a) and (b), respectively, for a better comparison with the data.}
\label{fig:ins_spinon}
\end{figure}

Both experiments and classical MC simulations reveal strong spin ice correlations in the system above \tn. Assuming the existence of a Coulombic phase with respect to $\tau^{\tilde x}$ above \tn, we have calculated the correlations based on the bosonic many-body theory of quantum spin ice \cite{SM,Hao2014}. As shown in Fig.~\ref{fig:ins_spinon}, the calculated dynamical structure factor at 450~mK (integrated over [0, 0.05]~meV) exhibits a pinch point pattern with additional broad scattering around [220] and [113] in good agreement with experimental data. According to the theory, besides the pinch point pattern at zero energy due to the scattering of the Coulombic phase, monopole creation and hopping cause broad scattering around [220] and [113]. Neutrons can be scattered by monopoles via two different processes: (i) the incoming neutron flips a spin belonging to an ice-rule tetrahedron creating a pair of monopoles, which gives a continuum at finite energy above a small gap; (ii) at finite temperature where there are a finite density of monopoles already in the system, the incoming neutron can flip a spin belonging to a monopole tetrahedron causing this monopole to hop which gives a continuum of scattering around zero energy. These scattering features are shown in Fig.~\ref{fig:ins_spinon}(b). This agrees with our data which exhibits strong broad signals around [220] and [113] at zero energy and a continuum at finite energy. However, the gapped feature is not clear in the data which could be attributed to possible fast decay of the coherent monopole excitations due to, for example, strong thermal fluctuations. The signal also may be contaminated by the scattering from possible short-range AIAO correlations. More experiments are needed to clarify this.

The existences of strong gapless spin-ice correlations above \tn\ and the gapped flat dynamical spin-ice modes in the ordered state \cite{Petit2016,Xu2019} make the ordering a candidate for a Higgs transition where the emergent gauge field of the Coulomb phase is gapped by the condensation of emergent gauge charges (monopoles) \cite{Chang2012,Powell2011}. Our semiclassical molecular dynamics calculations (Fig.~\ref{fig:mc_simu}(d) and Ref.~\cite{SM}) show that the gap closes with raising temperature which supports this picture. A Higgs transition was also proposed for pyrochlore Yb$_2$Ti$_2$O$_7$ by demonstrating the first order nature of the ordering transition and the sudden suppression of the intensity of the pinch point pattern below the transition \cite{Chang2012}. The gapped and gapless pinch point patterns below and above \tn\ is another important evidence for a Higgs transition.

Our results provide an explanation for the seemingly contradicting temperature dependences of the  intensities of the AIAO Bragg peak and the pinch point pattern in the energy-integrated polarized neutron scattering data reported for \ndzro\ \cite{Petit2016} and very recently for Nd$_2$ScNbO$_7$ \cite{Mauws2019}. It was shown that with increasing temperature, the magnetic Bragg peak weakens and disappears at \tn\ while the pinch point intensity maximizes around \tn\ and persists to much higher temperatures ($\sim1$~K) for both \ndzro\ and Nd$_2$ScNbO$_7$. This cannot be rationalized if the pinch point scattering is only a feature of the magnons in the ordered state. We argue that at \tn\ the gapped pinch point pattern is replaced by a pinch point pattern at zero energy due to the Coulombic phase built on ${\tilde\tau}^{\tilde x}$ which has the strongest ice correlations around \tn\ and gets weaker slowly with increasing temperature, similar to the temperature evolution of \avampx\ in the MC simulations [Fig.~\ref{fig:mc_simu}(b)]. The temperature range where the pinch point pattern presents is also comparable with $\tjx\sim1$~K. In addition, the Coulombic phase with strongly correlated spins could induce slow spin dynamics which supports the observed anomalously slow paramagnetic spin dynamics in the muon spin relaxation experiments \cite{Xu2016}.

What is more, it was reported recently that a spin ice model with frustrated transverse terms exhibits competing phases and nematicity \cite{Mathieu2017,Benton2018}. Our results provide a concrete experimental and theoretical example of an Ising antiferromagnet with frustrated transverse terms which also could show interesting physics. Our further MC simulations in Ref.~\cite{SM} show that the AIAO ordering temperature is suppressed largely with increasing $\tjx/|\tjz|$ which should be constant according to the mean field theory and on the other hand, the ordered phase invades the spin ice phase at finite temperature for $\tjx/|\tjz|>3$ similar to the phase diagram in Ref.~\cite{Mathieu2017} which is surprising because the spin ice phase should be more stable due to the higher entropy. This suggests that further theoretical study is needed.

In summary, we used high-energy-resolution inelastic neutron scattering, classical MC simulations, semi-classical molecular dynamics simulations and a bosonic theory of quantum spin ice to show that \ndzro\ has a non-trivial paramagnetic state with 2I2O spin-ice correlations, which is possibly a magnetic Coulombic phase, despite a long-range AIAO ordered ground state. We attributed it to the dominant frustrated transverse $\tjx$ term in the spin Hamiltonian and related the ordering transition to the Higgs mechanism. Our results indicate that the paramagnetic phase of an ordered system may host unconventional spin correlations different from the ground-state order in nature due to competition and frustration among different terms of anisotropic exchange interactions. This expands the field for searching for quantum spin ice to the ordered systems with frustrated terms in the spin Hamiltonian and makes it interesting to examine several similar QSI candidates with AIAO order, such as Nd$_2$(Hf/Sn/Pb)$_2$O$_7$, Nd$_2$ScNbO$_7$ and Sm$_2$(Ti/Sn)$_2$O$_7$ \cite{Anand2015,Bertin2015,Hallas2015,Mauws2018,Mauws2019,Viviane2019}.

\acknowledgements
We thank K. Siemensmeyer, Y.-P. Huang, M. Hermele, S. T. Bramwell and A. T. Boothroyd for helpful discussions on the related theory.  We acknowledge Helmholtz Gemeinschaft for funding via the Helmholtz Virtual Institute (Project No. VH-VI-521). This research used resources at the Spallation Neutron Source, a DOE Office of Science User Facility operated by the Oak Ridge National Laboratory. Experiments at the ISIS Neutron and Muon Source were supported by a beamtime allocation RB1810504 from the Science and Technology Facilities Council (DOI: 10.5286/ISIS.E.92924095).

\bibliographystyle{abbrv}

\renewcommand{\thefigure}{S\arabic{figure}}
\renewcommand{\thetable}{S\arabic{table}}

\renewcommand{\theequation}{S\arabic{equation}}

\makeatletter
\makeatother

\setcounter{figure}{0}
\setcounter{table}{0}
\setcounter{equation}{0}

\onecolumngrid
\newpage
\begin{center}
\textbf{\large Supplementary Information for \\
``Order out of a Coulomb phase and Higgs transtion: frustrated transverse interactions of Nd$_2$Zr$_2$O$_7$''}\\
J. Xu, Owen Benton, A. T. M. N. Islam, T. Guidi, G. Ehlers, B. Lake
\end{center}
\vspace{0.5cm}
\twocolumngrid

\section{\label{supp_sec:ins}INS data measured on CNCS and Osiris}
\begin{figure}[!htb]
\centering
\includegraphics[width=0.8\linewidth]{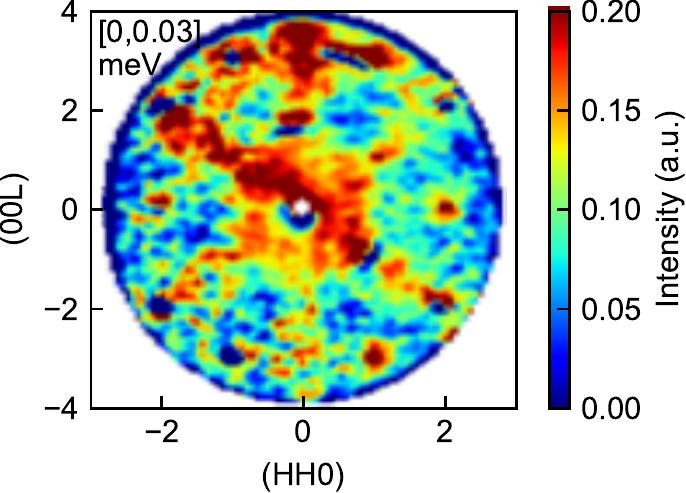}
\caption{Constant-energy slice of the raw data measured on CNCS integrated over [0, 0.03]~meV before averaging according to the symmetry of the (HHL) reciprocal plane.}
\label{fig:hhl_cncs_raw}
\end{figure}

Figure~\ref{fig:hhl_cncs_raw} shows constant energy slice of the (HHL) reciprocal plane (integrated over [0, 0.03]~meV) without symmetrization where a pinch point pattern is clearly shown.

\begin{figure}[!htb]
\centering
\includegraphics[width=\linewidth]{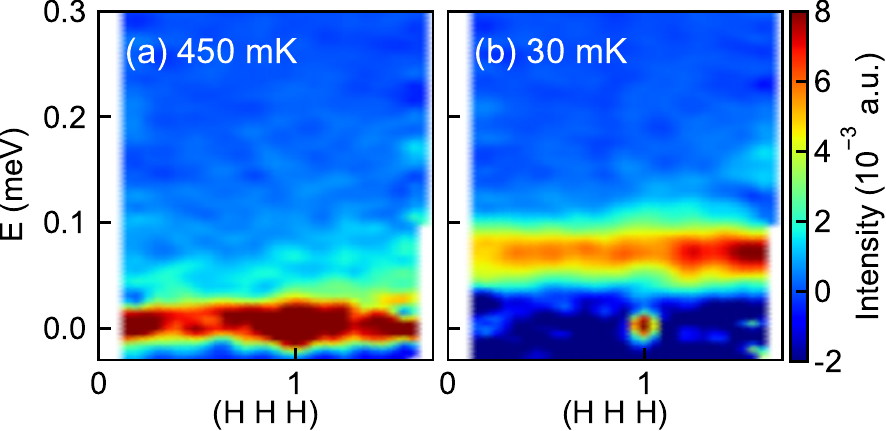}
\caption{$E-Q$ slices along (111) direction in reciprocal space of the Osiris data at 450~mK (a) and 30~mK (b).}
\label{fig:hhh_Osiris}
\end{figure}

Figure~\ref{fig:hhh_Osiris} shows the $E-Q$ slices along the (HHH) direction (an arm of the pinch point pattern) of the Osiris data measured at 450~mK and 30~mK which show that above \tn, the gapped pinch point pattern disappears and a pinch point pattern at zero energy appears.

\begin{figure*}[!htb]
\centering
\includegraphics[width=\textwidth]{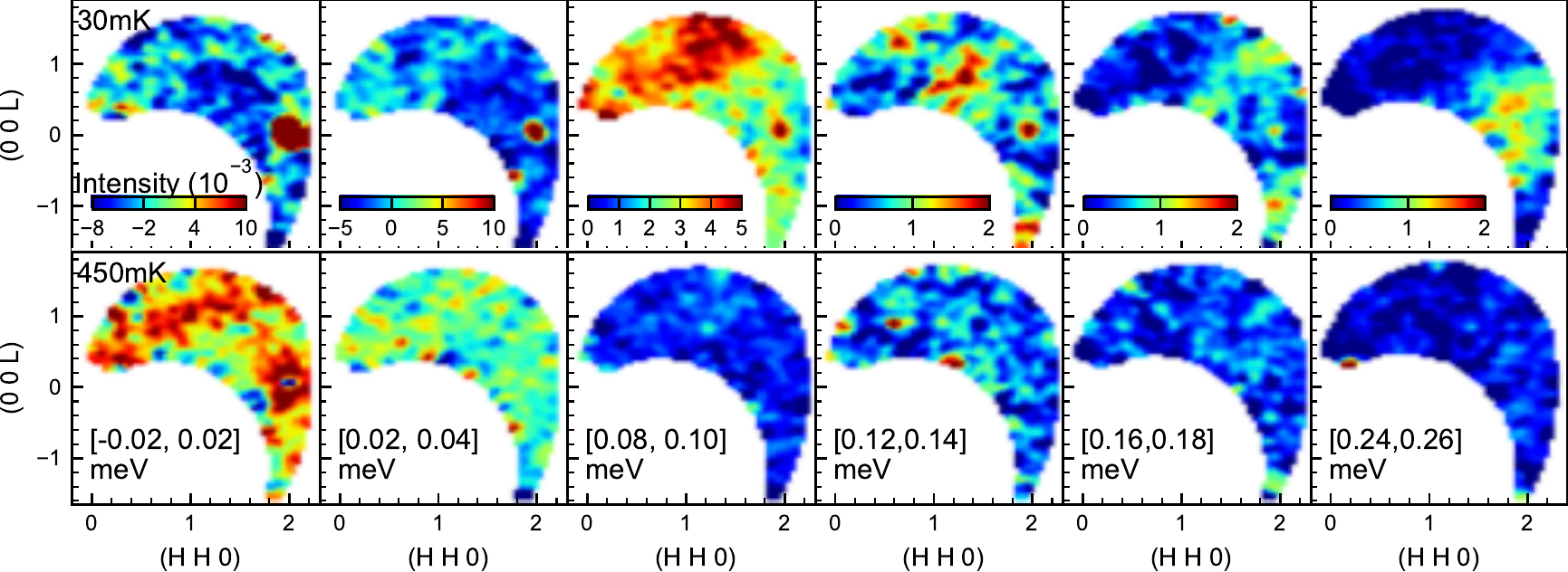}
\caption{Constant-energy slices in the (HHL) reciprocal plane of the INS data measured on Osiris comparing the 30~mK and 450~mK data. The strong sharp intensity at (220) at finite energy at 30~mK is an instrumental spurion resulted from leakage beyond the elastic channel.}
\label{fig:hhl_map_Osiris}
\end{figure*}

\begin{figure*}[!htb]
\centering
\includegraphics[width=0.9\textwidth]{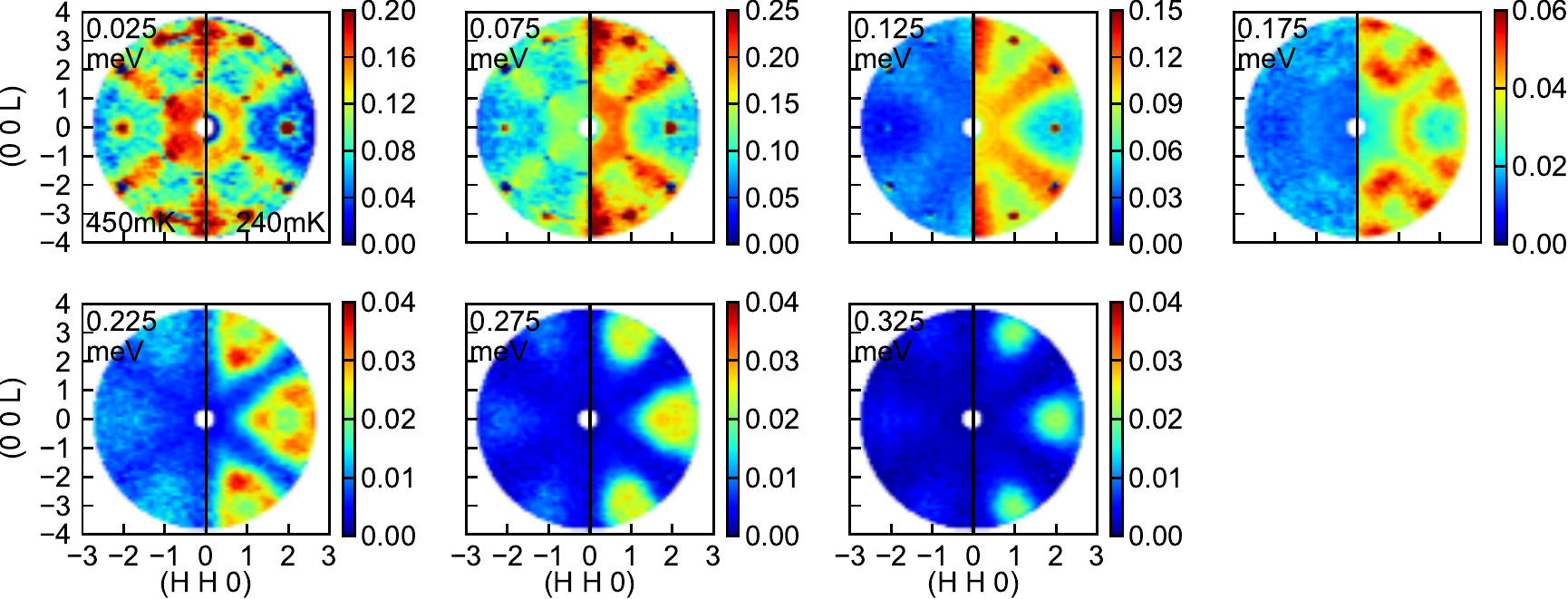}
\caption{Constant-energy slices of the (HHL) plane in the reciprocal space of the INS data measured on CNCS at 240~mK and 450~mK. The gapped pinch point pattern at 240~mK is replaced by a pinch point pattern at lower energy at 450~mK. However there are still weak magnon excitations at 450~mK which is associated to short-range AIAO correlations.}
\label{fig:hhl_map_cncs}
\end{figure*}

\begin{figure}[!htb]
\centering
\includegraphics[width=\linewidth]{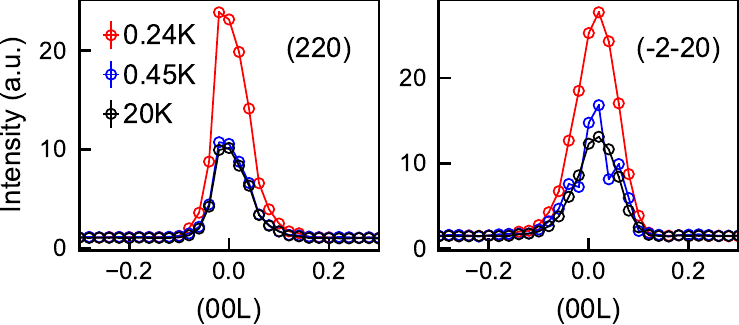}
\caption{Comparison of the (220) and (-2-20) peaks in the CNCS data at 240~mK, 450~mK and 20~K which shows that the long-range order at 240~mK and paramagnetic state at 450~mK.}
\label{fig:220_cncs}
\end{figure}

Fig.~\ref{fig:hhl_map_Osiris} shows the constant-energy slices in the (HHL) plane of the Osiris data measured at 30~mK and 450~mK. Fig.~\ref{fig:hhl_map_cncs} shows the constant-energy slices in the (HHL) plane of the CNCS data measured at 240~mK and 450~mK. Both show that the spinwave excitations disappeared or are extremely weak above \tn\ and a pinch point pattern at zero energy shows up. Fig.~\ref{fig:220_cncs} shows the disappearance of the magnetic Bragg intensity at (220) and (-2-20) at 450~mK in the CNCS data.

\section{\label{supp_sec:spinw}Quantum fluctuations in linear spin wave theory}
\begin{figure}[!htb]
\centering
\includegraphics[width=\linewidth]{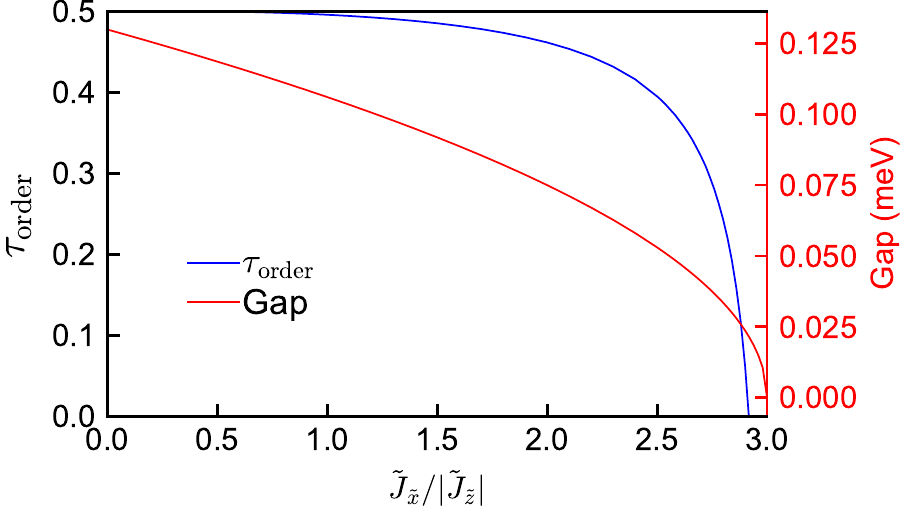}
\caption{Calculated ordered spin reduced by zero-point quantum reduction and gap to the flat magnon modes as functions $\tilde{J}_{\tilde x}$ based on the spin Hamiltonian for \ndzro\ in Ref.~\cite{Xu2019}.}
\label{fig:sw_mred_gap}
\end{figure}

Figure~\ref{fig:sw_mred_gap} shows the calculated reduced ordered spin for $0<\tjx/|\tjz|<3$ due to zero-point quantum fluctuations based linear spin wave theory using SPINW MATLAB package \cite{spinw}. The gap to flat modes are calculated from the equation \cite{Benton2016}
\begin{equation}
\text{gap} = \sqrt{(3|\tilde{J}_{\tilde{z}}| - \tilde{J}_{\tilde{x}})(3|\tilde{J}_{\tilde{z}}| -\tilde{J}_{\tilde{y}})}.
\label{eq:e_gap}
\end{equation}
They are calculated by changing $\tjx$ with fixing $\tjy$ and $\tjz$ of the spin Hamiltonian for \ndzro\ \cite{Xu2019}. With increasing $\tjx/|\tjz|$, the gap decreases and the quantum fluctuations are enhanced drastically for $\tjx/|\tjz|>2$ due to the proximity to a U(1) spin liquid. This can be related to the persistent spin dynamics in the muon spin relaxation experiments of \ndzro\ \cite{Xu2016}.

\section{\label{supp_sec:mc}Monte Carlo simulations}
The classical single-spin-flip Metropolis Monte Carlo simulations are performed using SPINW Matlab package by implementing our own codes \cite{spinw}. We used a $6\times6\times6$ supercell with periodic boundary conditions. At each temperature, $5\times10^4$ Monte Carlo steps per spin were used for thermal equilibrium and $5\times10^5-1\times10^6$ Monte Carlo steps per spin for data collection. To avoid freezing of the dynamics at low temperatures, the spins are moved in a cone around its present direction with the cone angle as a free parameter chosen to keep the acceptance rate above $30$\%.

\begin{figure}[!htb]
\centering
\includegraphics[width=0.8\linewidth]{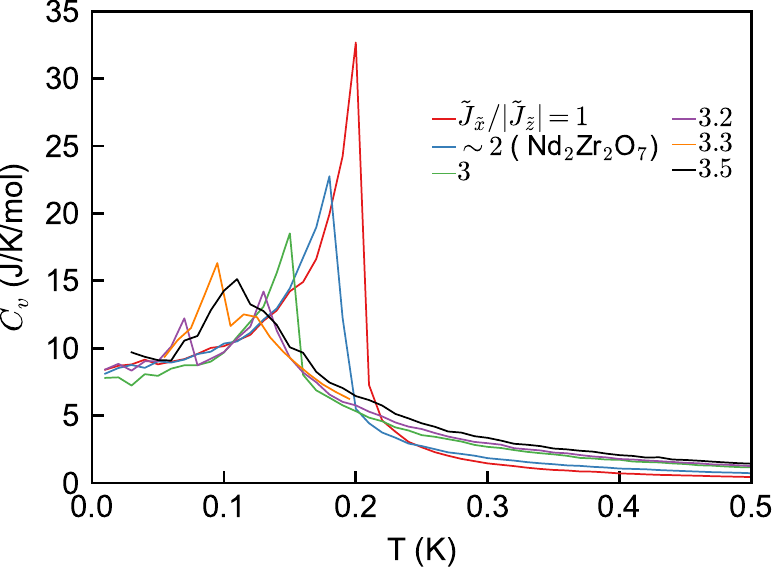}
\caption{Monte Carlo simulated specific heat as functions of temperature using the spin Hamiltonian of \ndzro\ in Ref.~\cite{Xu2019} with setting $\tjx=\{1,\sim2,3,3.2,3.3,3.5\}|\tjz|$. Up to $\tjx\sim3.1|\tjz|$, the system shows AIAO order but \tn\ is suppressed from $\sim0.20$\,K to $\sim0.13$\,K. There are two peaks for $\tjx/|\tjz|=3.2$ and 3.3. The peak at higher temperature is related to tendency of AIAO ordering of $\tau^{\tilde z}$ (see Fig.~\ref{fig:mc_tau_av}) and the one at lower temperature is associated with a re-entrant transition out of AIAO order to the spin ice regime. At $\tjx/|\tjz|=3.5 $ the system only shows crossover to spin ice state.}
\label{fig:mc_chi_cp}
\end{figure}

\begin{figure}[!htb]
\centering
\includegraphics[width=0.9\linewidth]{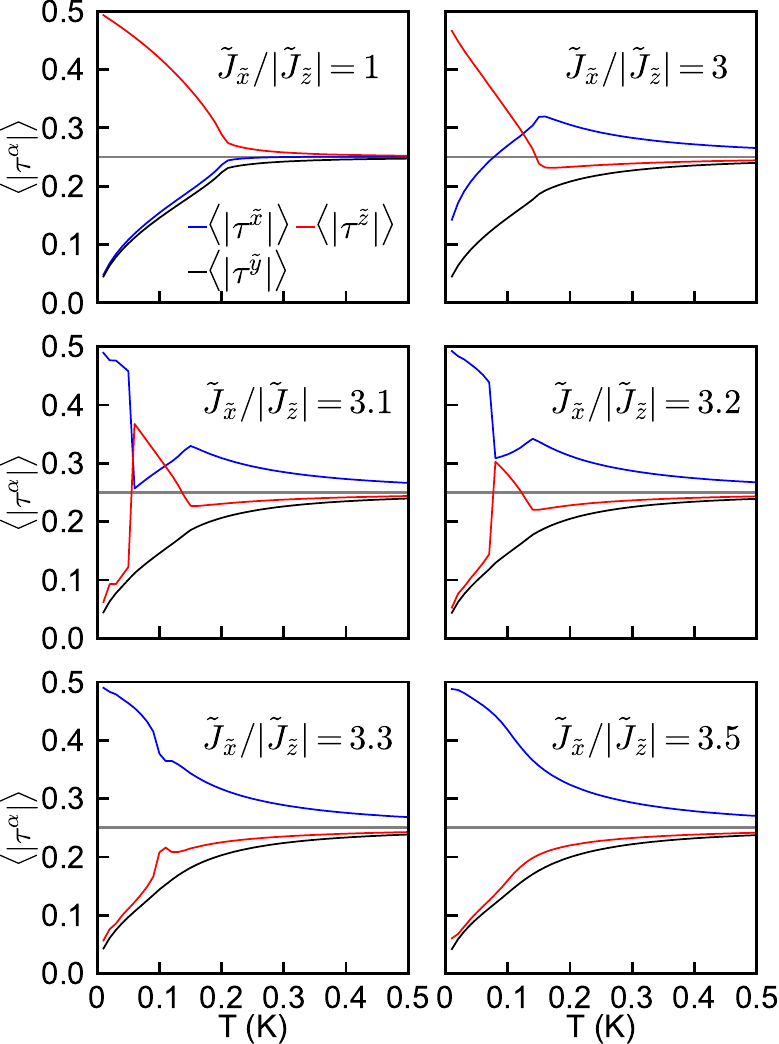}
\caption{Monte Carlo simulated $\langle|\tau^{\tilde\alpha}|\rangle$ ($\alpha=x,y,z$) using the spin Hamiltonian of \ndzro\ in Ref.~\cite{Xu2019} with setting $\tilde{J}_{\tilde x}=\{1,3,3.3,3.5\}|\tilde{J}_{\tilde z}|$. With increasing $\tjx/|\tjz|$, the system shows stronger and stronger spin-ice correlations above \tn\ which is suddenly suppressed at \tn. For $\tjx/|\tjz|=3.1, 3.3$, and $3.2$, the system shows tendency of AIAO ordering and then transition to spin ice state. For $\tjx/|\tjz|=3.5$, the system only shows crossover to spin ice state.}
\label{fig:mc_tau_av}
\end{figure}

Figure~\ref{fig:mc_chi_cp} shows the simulated temperature dependence of specific heat $C_{\text v}$ using the spin Hamiltonian of \ndzro\ with setting $\tilde{J}_{\tilde x}=\{1,\sim2,3,3.2,3.3,3.5\}|\tilde{J}_{\tilde z}|$ and Fig.~\ref{fig:mc_tau_av} shows $\langle|\tau^{\tilde\alpha}|\rangle$ ($\alpha=x,y,z$) as a function of temperature for different $\tjx/|\tjz|$. They show that spin-ice correlations become stronger above \tn\ with increasing $\tjx/|\tjz|$ and the AIAO ordering temperature is suppressed largely from $\sim0.20$~K for $\tjx/|\tjz|=1$ to $\sim0.13$~K for  $\tjx/|\tjz|=3$ which should be constant within the mean-field theory. It is very interesting that the system still shows a long-range AIAO order at the critical point $\tjx/|\tjz|=3$ where spin ice state is expected which has the same energy with the AIAO order but a lower free energy because of its higher entropy. For $\tjx/|\tjz|=3.2$ and $3.3$, $C_{\text v}$($T$) show double peaks which are related to tendency of AIAO ordering and a re-entrant transition out of AIAO order to the spin ice regime, as shown in Fig.~\ref{fig:mc_tau_av}. The interesting competition between the AIAO order and spin ice will be studies further.

\begin{figure*}[!htb]
\centering
\includegraphics[width=0.8\textwidth]{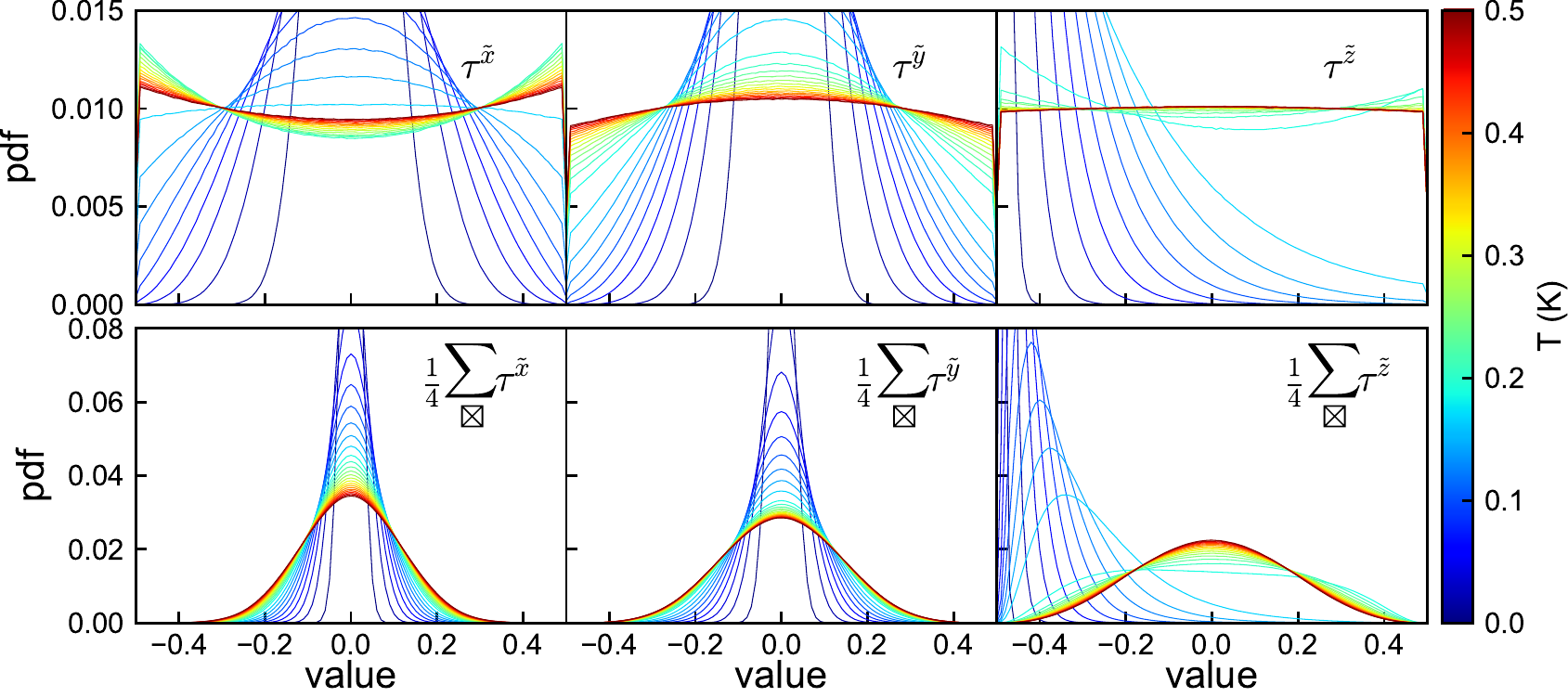}\\
\vspace{0.3cm}
\includegraphics[width=0.8\textwidth]{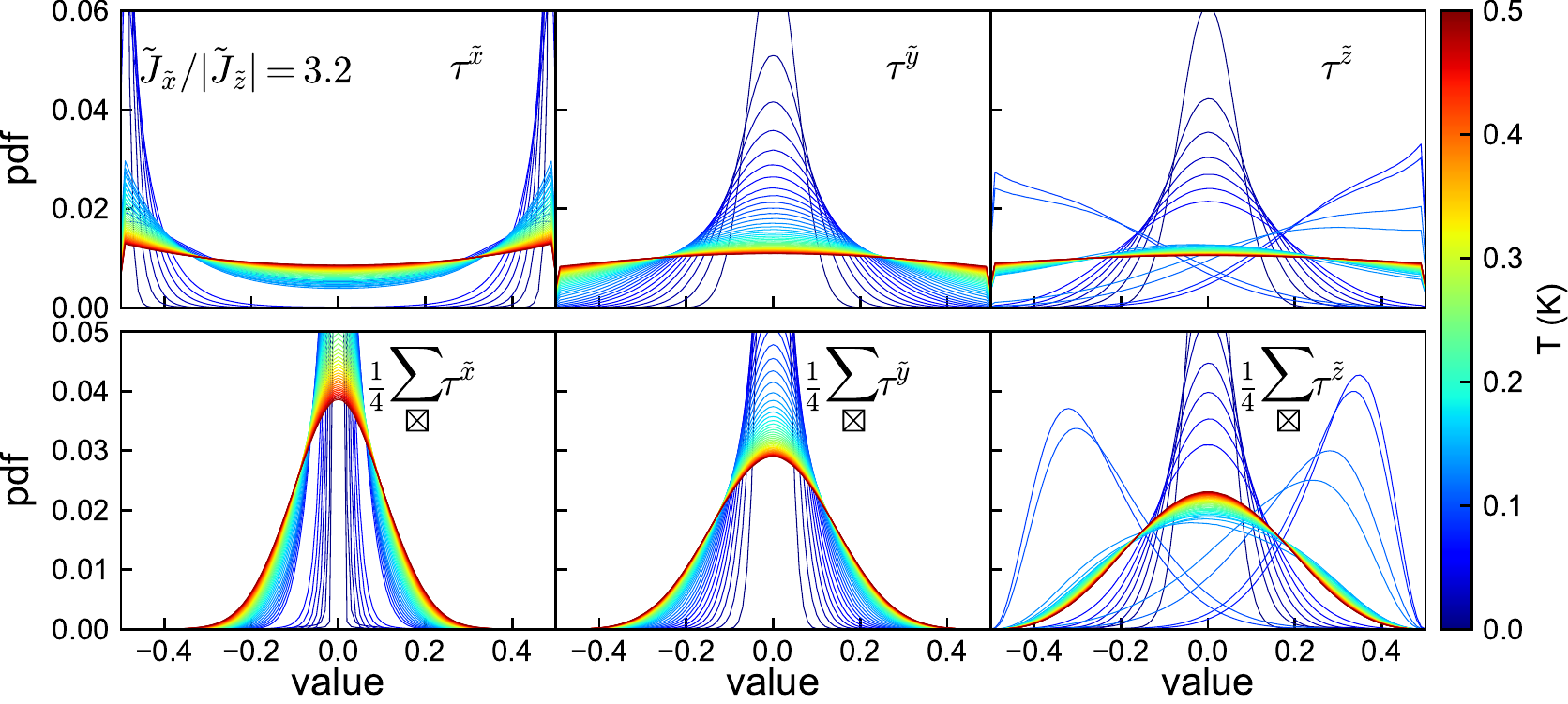}
\caption{Probability distribution functions of $\tau^{\tilde\alpha}$ and $1/4\sum_\boxtimes\tau^{\tilde\alpha}$ ($\alpha=x,y,z$) at different temperatures in Monte Carlo simulations using the spin Hamiltonian of \ndzro\ in Ref.~\cite{Xu2019} (first two rows) and $\tjx=3.2|\tjz|$ (last two rows). Note that the switch between the two types of AIAO domains leads to positive/negitve peak positions for $\tilde{\tau}^{\tilde z}$.}
\label{fig:mc_tau_hist}
\end{figure*}

Figure.~\ref{fig:mc_tau_hist} shows the probability distribution functions of $\tau^{\tilde\alpha}$ and the average over tetrahedra $1/4\sum_\boxtimes\tau^{\tilde\alpha}$ for $\tjx/|\tjz|\sim2$ (for \ndzro) and $3.2$. For \ndzro, strong spin-ice correlations of $\tau^{\tilde x}$ are built up with decreasing temperature and then suddenly suppressed below \tn. For $\tjx/|\tjz|=3.2$, the system first shows spin ice correlations at high temperature, then tendency of AIAO ordering and finally spin ice phase at low temperatures.

\section{\label{sec:mc_dyn}Spin dynamics at finite temperatures}
The spin dynamics at finite temperature were calculated using a semiclassical molecular dynamics. An ensemble of configurations is first obtained from Monte Carlo simulations with a supercell of $10\times10\times10$ cubic unit cells (16000 spins in total) with periodic boundary conditions. The ensemble is then evolved in time according to the equation of motion,
\begin{equation}
\frac{\textrm{d}\bm{\tau}_i(t)}{\textrm{d}t} = \bm{\tau}_i\times \sum_j J_{ij}\bm{\tau}_j(t)
\end{equation}
which is integrated numerically using eight-order Runge-Kutta method with an adaptive step-size control. In the simulation, $5\times10^5$ Monte Carlo steps were used for thermalization at a fixed temperature and measurements were taken every 1000 steps in $5\times10^5 \sim 2.5\times10^6$ Monte Carlo steps.

\begin{figure*}[!htb]
\centering
\includegraphics[width=0.8\textwidth]{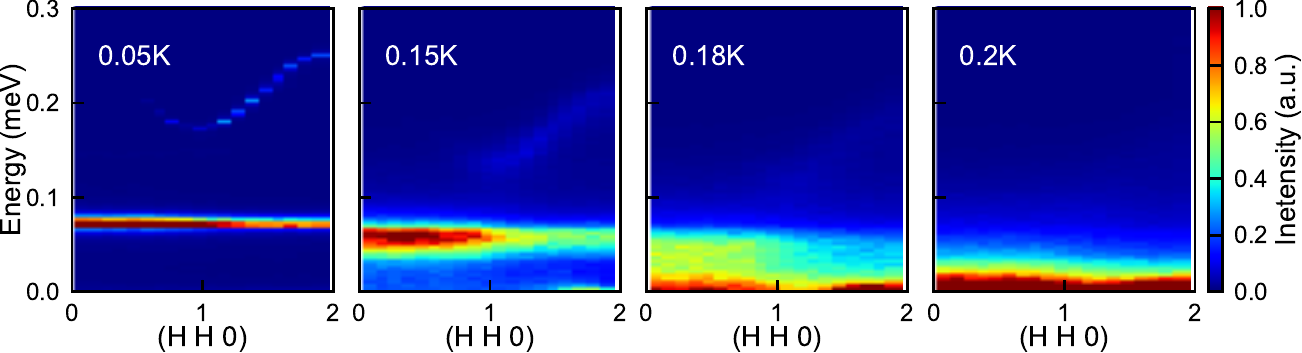}
\caption{Spin dynamics at finite temperatures simulated using semiclassical molecular dynamics showing the $E-Q$ slice along (HH0).}
\label{fig:mc_dyn_eq}
\end{figure*}

Fig.~\ref{fig:mc_dyn_eq} shows the simulated $E-Q$ slices along (HH0) direction at different temperatures which show the gap decreases with increasing temperature and most of the spectral weight changes to be elastic.

\section{\label{sec:theory}Theoretical calculation of structure factor in finite temperature quantum spin ice}

Here we give details of theoretical calculations of the structure factor $S(\mathbf{q}, \omega)$ for
$T>T_N$.
These calculations are made by modelling the system above $T_N$ as a spin ice with bosonic, propagating,
quantum monopole (spinon) excitations.
We first give some necessary definitions before discussing the low energy spin ice correlations [Sec.~\ref{subsec:gauge}]
and the non-interacting description of propagating monopoles  [Sec.~\ref{subsec:mon}].

The structure factor of interest is:
\begin{multline}
S(\mathbf{q}, \omega)=
\frac{1}{2\pi}\sum_{ij} \sum_{\mu \nu}\\
\int dt  \ e^{i \omega t}
\left( \delta_{\mu \nu}-\frac{q_{\mu} {q_{\nu} }}{q^2} \right)
\langle
m^{\mu}_i (\mathbf{q}, t) m^{\mu}_i (-\mathbf{q}, 0)
\rangle
\label{eq:sqomega}
\end{multline}
where $i,j=0,1,2,3$ index the four sublattices of the pyrochlore
lattice and $\mu, \nu$ are spatial directions in the crystal coordinate frame.
${\bf m}_i (\mathbf{q})$ is the Fourier transform of the magnetic moment
configuration on sublattice $i$:
\begin{multline}
{\bf m}_i (\mathbf{q})=  \\
\frac{1}{\sqrt{N_{\sf uc}}} \sum_{\bf R} \exp(-i \mathbf{q} \cdot (\mathbf{R}+ \mathbf{u}_i/2)) {\bf m}_i (\mathbf{R}+ \mathbf{u}_i/2) \
\label{eq:mqdef}
\end{multline}
where $N_{\sf uc}$ is the number of unit cells in the system, ${\bf R}$ are the centers of the unit cells, $\frac{1}{2}\mathbf{u}_i$
is the relative position of site $i$ within a unit cell and ${\bf m}_i (\mathbf{R}+ \mathbf{u}_i/2)$ is the magnetic moment on that site.
We take a coordinate frame where the vectors ${\bf u}_i$ are
\begin{eqnarray}
&&{\bf u}_0= (1,1,1), \
{\bf u}_1= (1,-1,-1), \nonumber \\
&&{\bf u}_2= (-1,1,-1), \
{\bf u}_3= (-1,-1,1).
\label{eq:udeff}
\end{eqnarray}

The relationship between the magnetic moments ${\bf m}_i$ and the pseudospin operators $\tau^{\tilde{\alpha}}_i$
which appear in the theoretical description of the magnetism \cite{Huang2014, Benton2016} is
\begin{eqnarray}
{\bf m}_i = g_{zz} \mu_B \left(  \cos(\vartheta) \tau^{\tilde{z}}_i + \sin(\vartheta) \tau^{\tilde{x}}_i \right) \hat{{\bf z}}_i \
\label{eq:m-S}
\end{eqnarray}
where $\mu_B$ is the Bohr magneton.
The parameters $g_{zz}$ and $\vartheta$ were estimated for Nd$_2$Zr$_2$O$_7$ in Ref.~\cite{Xu2019}:
\begin{eqnarray}
g_{zz}=5.0, \ \vartheta =0.98
\label{eq:g-theta}
\end{eqnarray}
and the local axes $\hat{\bf z}_i$ are
\begin{eqnarray}
\hat{\bf z}_i=\frac{1}{\sqrt{3}} {\bf u}_i.
\label{eq:local-axes}
\end{eqnarray}

Inserting Eq. (\ref{eq:m-S}) into Eq. (\ref{eq:sqomega}), and defining lattice Fourier transforms
of $\tau^{\tilde{\alpha}}_i$ analogously to Eq. (\ref{eq:mqdef}) gives:
\begin{multline}
S(\mathbf{q}, \omega)= g_{zz}^2 \mu_B^2
\frac{1}{2\pi}\sum_{ij}
 \left( \hat{\bf z}_i \cdot \hat{\bf z}_j - \frac{ (\hat{\bf z}_i \cdot \mathbf{q})(\hat{\bf z}_j \cdot \mathbf{q})}{q^2} \right)
\\
\int dt  \ e^{i \omega t}  \bigg( \cos^2(\vartheta) \langle \tau^{\tilde{z}}_i(\mathbf{q}, t)  \tau^{\tilde{z}}_i(-\mathbf{q}, 0) \rangle +
 \\
\cos(\vartheta) \sin(\vartheta) \langle \tau^{\tilde{z}}_i(\mathbf{q}, t)  \tau^{\tilde{x}}_i(-\mathbf{q}, 0) \rangle+
\\
\cos(\vartheta) \sin(\vartheta) \langle \tau^{\tilde{x}}_i(\mathbf{q}, t)  \tau^{\tilde{z}}_i(-\mathbf{q}, 0) \rangle +
\\
\sin^2(\vartheta) \langle \tau^{\tilde{x}}_i(\mathbf{q}, t)  \tau^{\tilde{x}}_i(-\mathbf{q}, 0) \rangle  \bigg).
\end{multline}

The symmetries of the XYZ Hamiltonian for dipolar-octupolar pyrochlores \cite{Huang2014} guarantee that the cross terms
$\langle \tau^{\tilde{z}} \tau^{\tilde{x}} \rangle$ vanish in the high-temperature phase.
Using this, we can express $S(\mathbf{q}, \omega)$ as a sum of two distinct contributions:
\begin{equation}
S(\mathbf{q}, \omega) = g_{zz}^2 \mu_B^2 \left(
\cos^2(\vartheta) S^{\tilde{z}\tilde{z}}(\mathbf{q}, \omega) +
\sin^2(\vartheta) S^{\tilde{x}\tilde{x}}(\mathbf{q}, \omega)
\right)
\label{eq:Sq-zz+xx}
\end{equation}
where
\begin{eqnarray}
&&S^{\tilde{z}\tilde{z}}(\mathbf{q}, \omega) =
\frac{1}{2\pi}\sum_{ij}
 \left( \hat{\bf z}_i \cdot \hat{\bf z}_j - \frac{ (\hat{\bf z}_i \cdot \mathbf{q})(\hat{\bf z}_j \cdot \mathbf{q})}{q^2} \right) \nonumber  \\
&& \qquad \int dt  \ e^{i \omega t}  \langle \tau^{\tilde{z}}_i(\mathbf{q}, t)  \tau^{\tilde{z}}_i(-\mathbf{q}, 0)
\label{eq:szz}
\rangle \\
&&S^{\tilde{x}\tilde{x}}(\mathbf{q}, \omega) =
\frac{1}{2\pi}\sum_{ij}
 \left( \hat{\bf z}_i \cdot \hat{\bf z}_j - \frac{ (\hat{\bf z}_i \cdot \mathbf{q})(\hat{\bf z}_j \cdot \mathbf{q})}{q^2} \right) \nonumber  \\
&& \qquad \int dt  \ e^{i \omega t}  \langle \tau^{\tilde{x}}_i(\mathbf{q}, t)  \tau^{\tilde{x}}_i(-\mathbf{q}, 0) \rangle
\label{eq:sxx}
\end{eqnarray}

Our model of the phase just above $T_N$ has spin ice correlations with respect to the $\tilde{x}$ axis of pseudospin space.
In this picture the correlator $S^{\tilde{x}\tilde{x}}(\mathbf{q}, \omega)$ measures processes within the ice manifold where
there are two spins with $\tau^{\tilde{x}} =\frac{1}{2}$ and two with $\tau^{\tilde{x}} =-\frac{1}{2}$ on each pyrochlore tetrahedron.
The correlator $S^{\tilde{z}\tilde{z}}(\mathbf{q}, \omega)$ then measures excitations out of that manifold:
i.e. monopoles.

The plot in Fig. 4 of the main text is made using Eq. (\ref{eq:Sq-zz+xx}), calculating $S^{\tilde{x}\tilde{x}}(\mathbf{q}, \omega)$
as described in Section \ref{subsec:gauge} and $S^{\tilde{z}\tilde{z}}(\mathbf{q}, \omega)$
as described in  Section \ref{subsec:mon}.

\subsection{\label{subsec:gauge}Low energy spin-ice correlations}

The calculation of  $S^{\tilde{x}\tilde{x}}(\mathbf{q}, \omega)$ assumes that at $T_N<T<\tilde{J}^{\tilde{x}}$,
the system fluctuates among states obeying an ice rule for $\tau^{\tilde{x}}$.

In a quantum spin ice there will be some energy scale, $g$, for quantum tunnelling between these states \cite{Hermele2004}.
At temperatures, $T\lesssim g$, the correlations will be significantly modified from their classical form with the signature
pinch point singularities being suppressed, and a photon dispersion being visible in the inelastic spectrum \cite{Benton2012}.
At higher temperatures $T>g$, the correlations of  $\tau^{\tilde{x}}$ will essentially take on their classical form \cite{Benton2012, Kato2015,Huang2018} and will be centred on $\omega=0$.
In our case, since to study the Coulomb phase we are restrictred to $T>T_N>g$, we assume the system to be in the latter regime where
the correlations of $\tau^{\tilde{x}}$  are essentially classical.

The structure factor of a classical spin ice is well known and has the form
\begin{equation}
S^{\tilde{x}\tilde{x}}(\mathbf{q}, \omega)=\frac{1}{2}\delta(\omega)
\sum_{ij}
 \left( \hat{\bf z}_i \cdot \hat{\bf z}_j - \frac{ (\hat{\bf z}_i \cdot \mathbf{q})(\hat{\bf z}_j \cdot \mathbf{q})}{q^2} \right) P_{ij}(\mathbf{q})
\label{eq:sq-spinice}
\end{equation}
where $P_{ij}(\mathbf{q})$ is the projection matrix given by Henley \cite{Henley2005}:
\begin{equation}
P_{ij}(\mathbf{q})=\delta_{ij}-\sum_{a=0,1, b=0,1}  E_{ia}(\mathbf{q}) M_{ab}(\mathbf{q}) E_{bj}^{\dagger}(\mathbf{q}).
\end{equation}

$E_{ia}(\mathbf{q})$ is a $4 \times 2$ matrix
\begin{eqnarray}
E(\mathbf{q})=\begin{pmatrix}
\exp(-i \mathbf{q} \cdot {\bf u}_0/2) & \exp(i \mathbf{q} \cdot {\bf u}_0/2)\\
\exp(-i \mathbf{q} \cdot {\bf u}_1/2) & \exp(i \mathbf{q} \cdot {\bf u}_1/2)\\
\exp(-i \mathbf{q} \cdot {\bf u}_2/2) & \exp(i \mathbf{q} \cdot {\bf u}_2/2)\\
\exp(-i \mathbf{q} \cdot {\bf u}_3/2) & \exp(i \mathbf{q} \cdot {\bf u}_3/2)
\end{pmatrix}
\end{eqnarray}
and $M_{ab} (\mathbf{q})$ is a $2 \times 2$ matrix
\begin{eqnarray}
M(\mathbf{q})=(E^{\dagger} (\mathbf{q}) \cdot E(\mathbf{q}))^{-1}.
\end{eqnarray}

To account for finite energy resolution in the experiment, we
replace the delta function in Eq. (\ref{eq:sq-spinice}) with a Gaussian:
\begin{eqnarray}
\delta(\omega) \to G(\omega)=\frac{2}{f}\sqrt{\frac{\log(2)}{\pi}} e^{-\frac{4 \log(2)\omega^2}{f^2}}
\label{eq:gaussian}
\end{eqnarray}
with full width at half maximum (FWHM) $f=0.1$meV for Fig.~\ref{fig:ins_spinon}(a) and $f=0.02$meV for Fig.~\ref{fig:ins_spinon}(b).

\subsection{\label{subsec:mon}Non-interacting model for propagating monopoles}

The calculation of $S^{\tilde{z}\tilde{z}}$ is based on the intuition
that if the system obeys an ice rule with respect to $\tau^{\tilde{x}}$, the
action of $\tau^{\tilde{z}}$ will create pairs of violations of the ice rule: namely, monopoles.
Our treatment of the monopoles follows the non-interacting theory introduced in
Ref. \cite{Hao2014}.

Introducing ladder operators with respect to the $\tilde{x}$ axis of spin space:
\begin{eqnarray}
\tau^{\tilde{\pm}}_i=\tau^{\tilde{y}}_i \pm i \tau_i^{\tilde{z}}
\end{eqnarray}
we can write the Hamiltonian as
\begin{eqnarray}
&&\mathcal{H}=
\sum_{\langle ij \rangle}
\bigg[
\tilde{J}^{\tilde{x}}
\tau^{\tilde{x}}_i \tau^{\tilde{x}}_j  -
\tilde{J}^{\tilde{\pm}} \left(  \tau^{\tilde{+}}_i \tau^{\tilde{-}}_j+
 \tau^{\tilde{-}}_i \tau^{\tilde{+}}_j  \right) + \nonumber \\
&& \qquad \qquad
\tilde{J}^{\tilde{\pm} \tilde{\pm}} \left(  \tau^{\tilde{+}}_i \tau^{\tilde{+}}_j+
 \tau^{\tilde{-}}_i \tau^{\tilde{-}}_j  \right)
\bigg]
\label{eq:H-ladder}
\end{eqnarray}
In Nd$_2$Zr$_2$O$_7$ we have \cite{Xu2019} $\tilde{J}^{\tilde{x}}\approx0.091$meV,
$\tilde{J}^{\tilde{\pm}}=-\frac{1}{4} \left(\tilde{J}^{\tilde{y}} + \tilde{J}^{\tilde{z}}\right) \approx 0.008$meV,
$\tilde{J}^{\tilde{\pm} \tilde{\pm}}=\frac{1}{4} \left(\tilde{J}^{\tilde{y}} - \tilde{J}^{\tilde{z}}\right) \approx 0.015$meV.

The $\tilde{J}^{\tilde{x}}$ term in Eq. (\ref{eq:H-ladder}) exerts an energy cost $\tilde{J}^{\tilde{x}}/2$ per monopole.
The $\tilde{J}^{\tilde{\pm}}$ term generates monopole hopping.
$\tilde{J}^{\tilde{\pm}\tilde{\pm}}$ is an interaction term for the monopoles.

The monopoles are defined on the pyrochlore tetrahedra. The centers of these
tetrahedra form a diamond lattice, which may be divided into two interpenetrating
FCC sublattices A and B.
The monopole charge at a tetrahedron ${\bf R}$ is given by
\begin{eqnarray}
Q_{\bf R}= \epsilon_{\bf R} \sum_{i=0}^{3} \tau^{\tilde{x}}_i ({\bf R}+{\bf u}_i/2)
\end{eqnarray}
where $ \epsilon_{\bf R}=+1$ for `A' sublattice tetrahedra and
$ \epsilon_{\bf R}=-1$ for `A' sublattice tetrahedra.

Following Ref. \cite{Hao2014}, we can relate the spin ladder operators
to raising and lowering operators of the charge $\psi^{\dagger}_{\bf R}, \psi_{\bf R}$
\begin{eqnarray}
&&  \tau^{\tilde{+}}_i ({\bf R}+{\bf u}_i/2) = \frac{1}{2} \psi^{\dagger}_{\bf R}  \exp(i A_{{\bf R},{\bf R}+{\bf u}_i})
\psi_{{\bf R}+{\bf u}_i} \quad
 \\
&& \tau^{\tilde{-}}_i ({\bf R}+{\bf u}_i/2) = \frac{1}{2} \psi_{\bf R}  \exp(-i A_{{\bf R},{\bf R}+{\bf u}_i})
\psi^{\dagger}_{{\bf R}+{\bf u}_i}. \quad
\end{eqnarray}
$A_{{\bf R} {\bf R}'}$ is the gauge field to which the charges are coupled.
In the spirit of the mean field approximation of
Ref. \cite{Savary2012} we decouple the gauge field from the
monopoles and replace
\begin{eqnarray}
\frac{1}{2} \langle \exp(i A_{{\bf R},{\bf R}+{\bf u}_i})  \rangle \to \kappa.
\end{eqnarray}
to write
\begin{eqnarray}
&&  \tau^{\tilde{+}}_i ({\bf R}+{\bf u}_i/2) = \kappa \psi^{\dagger}_{\bf R}
\psi_{{\bf R}+{\bf u}_i} \quad
\label{eq:s+charge}
 \\
&& \tau^{\tilde{-}}_i ({\bf R}+{\bf u}_i/2) = \kappa \psi_{\bf R}
\psi^{\dagger}_{{\bf R}+{\bf u}_i}. \quad
\label{eq:s-charge}
\end{eqnarray}

For our present purposes we will determine $\kappa$ later from the sum rule on the structure factor.
The charge raising and lowering operators obey a constraint
\begin{eqnarray}
\psi_{\bf R}^{\dagger} \psi_{\bf R}=1.
\end{eqnarray}

As in Ref. \cite{Hao2014} we write $\psi_{\bf R}$, $Q_{\bf R}$ in terms of boson operators
$d_{\bf R}, b_{\bf R}$ which respectively carry postive and negative charge.
\begin{eqnarray}
&&\psi_{\bf R}=\frac{1}{\sqrt{1+d_{\bf R}^{\dagger} d_{\bf R}
+b_{\bf R}^{\dagger} b_{\bf R}
}} ( d_{\bf R}+b_{\bf R}^{\dagger} )
\label{eq:psi-ab}
\\
&& Q_{\bf R}=d_{\bf R}^{\dagger} d_{\bf R}-b_{\bf R}^{\dagger} b_{\bf R}.
\end{eqnarray}
The $d$ and $b$ bosons obey a  constraint, that they
cannot occupy the same site at the same time
\begin{eqnarray}
\left( d_{\bf R}^{\dagger} d_{\bf R}   \right)\left( b_{\bf R}^{\dagger} b_{\bf R}   \right)=0.
\end{eqnarray}

Assuming low-density of the bosons we can also expand
the denominator of Eq. (\ref{eq:psi-ab}) to write
\begin{eqnarray}
&&\psi_{\bf R}\approx d_{\bf R}+b_{\bf R}^{\dagger} \\
&&\psi_{\bf R}^{\dagger} \approx d_{\bf R}^{\dagger}+b_{\bf R}
\end{eqnarray}

We can then write $\mathcal{H}_\tau$ in terms of spinon operators, keeping only terms
up to bilinear order:
\begin{multline}
\mathcal{H}_\tau\approx
\sum_{{\bf R} \in A, B}
\bigg\{
\frac{{ J}^{\tilde{x}}}{2}
(d^{\dagger}_{{\bf R}}
d_{{\bf R}} +
b^{\dagger}_{{\bf R}}
b_{{\bf R}}
)  - \kappa^2 \tilde{J}^{\tilde{\pm}}
\sum_{i<j}
\\
 \bigg[
(d_{{\bf R}+ \epsilon_{\bf R} {\bf u}_i}+ b^{\dagger}_{{\bf R}+ \epsilon_{\bf R} {\bf u}_i})
(d^{\dagger}_{{\bf R}+ \epsilon_{\bf R} {\bf u}_j}+ b_{{\bf R}+  \epsilon_{\bf R} {\bf u}_j})
\\
+
(d^{\dagger}_{{\bf R}+ \epsilon_{\bf R} {\bf u}_i}+ b_{{\bf R}+  \epsilon_{\bf R} {\bf u}_i})
(d_{{\bf R}+ \epsilon_{\bf R} {\bf u}_j}+ b^{\dagger}_{{\bf R}+  \epsilon_{\bf R} {\bf u}_j})
\bigg]
\bigg\}.
\label{eq:H-bd}
\end{multline}

The first term in Eq. (\ref{eq:H-bd})
is simply a chemical potential term for the
spinons while the second incorporates both hopping and
charge conserving
pair creation/annihilation processes.
The coefficients of these terms may be renormalized from their
bare values by both fluctuations of the gauge field and by interactions.
To take this into account we will take
\begin{eqnarray}
\frac{J^{\tilde{x}}}{2}  \to \mu; \qquad
\kappa^2 {J}^{\tilde{\pm}}  \to \eta;
\label{eq:phenom}
\end{eqnarray}
and consider $\mu$ and $\eta$ as phenomenological
parameters.

We are then left with two copies (one on the $A$ sublattice, one on the $B$ sublattice)
of a non-interacting boson Hamiltonian on the FCC lattice.

To diagonalize Eq. (\ref{eq:H-bd}) we Fourier transform to reciprocal space,
and then do a Bogoliubov  transformation
\begin{eqnarray}
&&d_{\mathbf{q} \alpha}=\cosh(\theta_{\mathbf{q}}) \tilde{d}_{\mathbf{q} \alpha}
+ \sinh(\theta_{\mathbf{q}}) \tilde{b}^{\dagger}_{-\mathbf{q} \alpha}
\label{eq:d-bogoliubov}
\\
&&b_{\mathbf{q} \alpha}=\cosh(\theta_{\mathbf{q}}) \tilde{b}_{\mathbf{q} \alpha}
+ \sinh(\theta_{\mathbf{q}}) \tilde{d}^{\dagger}_{-\mathbf{q} \alpha}
\label{eq:b-bogoliubov}
\end{eqnarray}
where
\begin{eqnarray}
&&\gamma(\mathbf{q})=
2 \bigg(\cos{\left( 2q_x\right)} \cos{\left(
 2q_y \right) }
+\cos{\left( 2 q_x \right)} \cos{\left(
 2q_z \right) }
 \nonumber \\
&&
\qquad
+\cos{\left(2q_y \right)} \cos{\left(
 2q_z \right) }
\bigg) \\
&&\sinh(2 \theta_{\mathbf{q}})= \frac{2\eta \gamma(\mathbf{q})}{\omega(\mathbf{q})}
\label{eq:sinh2theta}\\
&&\cosh(2 \theta_{\mathbf{q}})= \frac{\mu- 2 \eta \gamma(\mathbf{q})}{\omega(\mathbf{q})}
\label{eq:cosh2theta}\\
&&\omega_{\mathbf{q}}= \mu \sqrt{1-\frac{4 \eta \gamma(\mathbf{q})}{\mu}}
\label{eq:dispersion}.
\end{eqnarray}

This brings the Hamiltonian into the form
\begin{eqnarray}
&&H_\tau\approx\sum_{\alpha=A, B}\sum_{\mathbf{q}} \bigg[ \omega_{\mathbf{q}}
\left(
\tilde{d}^{\dagger}_{\mathbf{q}, \alpha} \tilde{d}_{\mathbf{q}, \alpha}
+\tilde{b}^{\dagger}_{\mathbf{q}, \alpha} \tilde{b}_{\mathbf{q}, \alpha}
 \right)+  \nonumber \\
&&\qquad \omega_{\mathbf{q}}-(\mu-2 \eta \gamma(\mathbf{q}))
\bigg]
\label{eq:Hspinon-diag}
\end{eqnarray}
with the spinon dispersion  $\omega_{\mathbf{q}}$
given by Eq. (\ref{eq:dispersion}).

We then seek to turn the non-interacting treatment
of the spinons into a calculation of $S^{\tilde{z}\tilde{z}}(\mathbf{q}, \omega)$.
We first write
\begin{multline}
\langle {\tau}^{\tilde{z}}_i(\mathbf{q}, t)
{\tau}^{\tilde{z}}_j(-\mathbf{q}, 0) \rangle=
\\
\frac{1}{4}
\big(
\langle {\tau}^{\tilde{-}}_i(\mathbf{q}, t)
{\tau}^{\tilde{+}}_j(-\mathbf{q}, 0) \rangle
+
\langle {\tau}^{\tilde{+}}_i(\mathbf{q}, t)
{\tau}^{\tilde{-}}_j(-\mathbf{q}, 0) \rangle
\\
-
\langle {\tau}^{\tilde{+}}_i(\mathbf{q}, t)
{\tau}^{\tilde{+}}_j(-\mathbf{q}, 0) \rangle
-
\langle {\tau}^{\tilde{-}}_i(\mathbf{q}, t)
{\tau}^{\tilde{-}}_j(-\mathbf{q}, 0) \rangle
\big)
\label{eq:Sxx-Spm-Spmpm}
\end{multline}
and note that the spinon hopping Hamiltonian
conserves the total spinon charge on each FCC sublattice independently.
The correlators $\langle {\tau}^{+}_i(-\mathbf{q}, t)
{\tau}^{+}_j(\mathbf{q}, 0) \rangle$
and $\langle {\tau}^{-}_i(-\mathbf{q}, t)
{\tau}^{-}_j(\mathbf{q}, 0) \rangle$ appearing in Eq. (\ref{eq:Sxx-Spm-Spmpm})
do not conserve these charges
independently and must therefore vanish within this approximation.
We are therefore left with
\begin{multline}
\langle {\tau}^{\tilde{z}}_i(\mathbf{q}, t)
{\tau}^{\tilde{z}}_j(-\mathbf{q}, 0) \rangle=
\\
\frac{1}{4}
\big(
\langle {\tau}^{\tilde{-}}_i(\mathbf{q}, t)
{\tau}^{\tilde{+}}_j(-\mathbf{q}, 0) \rangle+
\langle {\tau}^{\tilde{+}}_i(\mathbf{q}, t)
{\tau}^{\tilde{-}}_j(-\mathbf{q}, 0) \rangle
\label{eq:Sxx-Spm}
\end{multline}

The spin operators relate to the charge raising and lowering
operators through Eqs. (\ref{eq:s+charge})- (\ref{eq:s-charge}).
Using these relationships,  leads to the correlation function of interest:
\begin{eqnarray}
&&
\langle \tau^{{z}}_i(\mathbf{q}, t)
\tau^{{z}}_j(-\mathbf{q}, 0) \rangle=
\frac{\kappa^2}{4 N_{\sf u.c.}}
\sum_{\mathbf{k}} \sum_{\mathbf{k}'}
 \nonumber \\
&&
\bigg[
e^{-i \mathbf{u}_i \cdot (\mathbf{k}+\mathbf{q}/2 )}
e^{i \mathbf{u}_j \cdot (\mathbf{k}'+\mathbf{q}/2 )}
 \nonumber \\
&&
\langle
\psi^{\dagger}_{\mathbf{q}+\mathbf{k}{A}}(t)
\psi^{\phantom\dagger}_{\mathbf{k} B}(t)
\psi^{\phantom\dagger}_{\mathbf{q}+\mathbf{k}'{A}}(0)
\psi^{\dagger}_{\mathbf{k}' B}(0)
\rangle
+
 \nonumber \\
&&
e^{i \mathbf{u}_i \cdot (\mathbf{k}'-\mathbf{q}/2 )}
e^{-i \mathbf{u}_j \cdot (\mathbf{k}-\mathbf{q}/2 )}
 \nonumber \\
&&
\langle
\psi^{\phantom\dagger}_{-\mathbf{q}+\mathbf{k}{A}}(t)
\psi^{\dagger}_{\mathbf{k} B}(t)
\psi^{\dagger}_{-\mathbf{q}+\mathbf{k}'{A}}(0)
\psi^{\phantom\dagger}_{\mathbf{k}' B}(0) \rangle
\bigg]
\label{eq:szsz-psi}
\end{eqnarray}

The expectation values  in Eq. (\ref{eq:szsz-psi}) can be calculated from the
non-interacting theory, as products of bilinear expectations values:
\begin{eqnarray}
&&\langle \psi_{\mathbf{q} \alpha} (t)  \psi^{\dagger}_{\mathbf{q}' \alpha} (0) \rangle=
\langle \psi^{\dagger}_{\mathbf{q} \alpha} (t)  \psi_{\mathbf{q}' \alpha} (0) \rangle= \nonumber\\
&& \frac{\mu}{\omega(\mathbf{q})} \delta(\mathbf{q}-\mathbf{q}')
\left[
(1+n_B(\omega(\mathbf{q}))) e^{-i \omega_{\mathbf{q}} t} +n_B(\omega(\mathbf{q}))e^{i \omega_{\mathbf{q}} t}
\right]. \nonumber \\
\end{eqnarray}
$n_B(\omega)$ is the Bose-Einstein distribution at temperature $T$.

Using these we obtain the final result for $S^{\tilde{z} \tilde{z}}(\mathbf{q}, \omega)$:
\begin{eqnarray}
&&S^{\tilde{z} \tilde{z}}(\mathbf{q}, \omega)=\frac{\kappa^2}{2 N_{\sf u.c.}}
\sum_{\mathbf{k}} \frac{\mu^2 \Gamma(\mathbf{q}, \mathbf{k})}{\omega(\mathbf{k}) \omega(\mathbf{q}+\mathbf{k}) } \bigg( \nonumber \\
&& \quad\quad
\delta(\omega-\omega(\mathbf{k})-\omega(\mathbf{q}+\mathbf{k})) \times \nonumber \\
&&\quad\quad
(1+n_B(\omega(\mathbf{k})))(1+n_B(\omega(\mathbf{q}+\mathbf{k})))
+
\nonumber \\
&&\quad\quad
\delta(\omega-\omega(\mathbf{k})+\omega(\mathbf{q}+\mathbf{k})) \times \nonumber \\
&&\quad\quad
 (1+n_B(\omega(\mathbf{k})))n_B(\omega(\mathbf{q}+\mathbf{k}))+
\nonumber \\
&&\quad\quad
\delta(\omega+\omega(\mathbf{k})-\omega(\mathbf{q}+\mathbf{k}))   \times \nonumber \\
&&\quad\quad
n_B(\omega(\mathbf{k})) (1+n_B(\omega(\mathbf{q}+\mathbf{k})))+
\nonumber \\
&&\quad\quad
\delta(\omega+\omega(\mathbf{k})+\omega(\mathbf{q}+\mathbf{k})) \times \nonumber \\
&&\quad\quad
   n_B(\omega(\mathbf{k}))  n_B(\omega(\mathbf{q}+\mathbf{k}))
 \bigg)
\label{eq:szsz-final}
\end{eqnarray}
where
\begin{equation}
\Gamma(\mathbf{q}, \mathbf{k})=\sum_{i,j} \cos((\mathbf{k}+\mathbf{q}/2) \cdot (\mathbf{u}_i - \mathbf{u}_j)) \left(  \delta_{ij} - \frac{q_i q_j}{q^2} \right).
\end{equation}

The first term in Eq. (\ref{eq:szsz-final}) comes from the creation of monopoles pairs by an incoming neutron,
the final one from the annihilation of monopole pairs and the middle two from processes where a neutron
flips a spin to hop an already-present monopole.

For the purposes of the plot in Fig. 4 of the manuscript Eq. (\ref{eq:szsz-final}) is evaluated with Monte Carlo integration over the Brillouin zone for
each value of ${\bf q}$. In the calculations we set $\mu=0.05$meV,  $\eta=0.002$meV, $T=450$mK. $\kappa$ is fixed by requiring agreement with the sum rule:
\begin{eqnarray}
\sum_{\mathbf{q}} \langle \tau_i^{\tilde{z}} (\mathbf{q}, t=0) \tau_i^{\tilde{z}} (-\mathbf{q}, t=0)
\rangle=\frac{N_{\sf uc}}{4}
\end{eqnarray}

Finite energy resolution is accounted for by replacing the delta functions in Eq. (\ref{eq:szsz-final})
with Gaussians of FWHM $f=0.1$meV for Fig.~\ref{fig:ins_spinon}(a) and $f=0.02$meV for Fig.~\ref{fig:ins_spinon}(b) [Eq. (\ref{eq:gaussian})].


\begin{thebibliography}{10}

\bibitem{Lacroix2011book}
C. Lacroix, P. Mendels, and F. Mila, {\it Introduction to frustrated magnetism: materials, experiments, theory} (Springer Science and Business Media, 2011), Vol. 164.

\bibitem{Gardner2010rev}
J. S. Gardner, M. J. P. Gingras, and J. E. Greedan, Rev. Mod. Phys. {\bf 82}, 53 (2010).

\bibitem{Fennell2009}
T. Fennell, P. P. Deen, A. R. Wildes, K. Schmalzl, D. Prabhakaran, A. T. Boothroyd, R. J. Aldus, D. F. McMorrow, and S. T. Bramwell, Science {\bf 326}, 415  (2009).

\bibitem{Morris2009}
D. J. P. Morris, D. A. Tennant, S. A. Grigera, B. Klemke, C. Castelnovo, R. Moessner, C. Czternasty, M. Meissner, K. C. Rule, J.-U. Hoffmann, K. Kiefer, S. Gerischer, D. Slobinsky, R. S. Perry, Science {\bf 326}, 411 (2009).

\bibitem{Harris1997}
M. J. Harris, S. T. Bramwell, D. F. McMorrow, T. Zeiske, and K. W. Godfrey, Phys. Rev. Lett. {\bf 79}, 2554 (1997).

\bibitem{Henley2005}
C. L. Henley, Phys. Rev. B {\bf 71}, 014424 (2005).

\bibitem{Hermele2004}
M. Hermele, Matthew P. A. Fisher, and L. Balents, Phys. Rev. B {\bf 69}, 064404 (2004).

\bibitem{Shannon2012}
Nic Shannon, Olga Sikora, Frank Pollmann, Karlo Penc, and Peter Fulde, Phys. Rev. Lett. {\bf 108} 067204 (2012).

\bibitem{Savary2012}
L. Savary and L. Balents, Phys. Rev. Lett. {\bf 108}, 037202 (2012).

\bibitem{Onoda2010}
S. Onoda and Y. Tanaka, Phys. Rev. Lett. {\bf 105}, 047201 (2010).

\bibitem{Onoda2011}
S. Onoda and Y. Tanaka, Phys. Rev. B {\bf 83}, 094411 (2011).

\bibitem{Benton2012}
O. Benton, O. Sikora and N. Shannon,
Phys. Rev. B {\bf 86}, 075154 (2012)

\bibitem{Lee2012}
S. B. Lee, S. Onoda, and L. Balents, Phys Rev B {\bf 86}, 104412 (2012).

\bibitem{Kato2015}
Y. Kato and S. Onoda, Phys. Rev. Lett. {\bf 115}, 077202 (2015).

\bibitem{Huang2018}
C. J. Huang, Y. Deng, Y. Wan and Z. Y. Meng, Phys. Rev. Lett. {\bf 120}, 167202, (2018).

\bibitem{Gingras2014}
M. J. P. Gingras and P. A. McClarty, Rep. Prog. Phys. {\bf 77}, 056501 (2014).

\bibitem{Ross2011}
Kate A. Ross, Lucile Savary, Bruce D. Gaulin, and Leon Balents, Phys. Rev. X {\bf 1}, 021002 (2011).

\bibitem{Kimura2013}
K. Kimura, S. Nakatsuji, J-J. Wen, C. Broholm, M. B. Stone, E. Nishibori and H. Sawa, Nat. Commu. {\bf 4}, 1934 (2013).

\bibitem{Sibille2018}
R. Sibille, N. Gauthier, Han Yan, Monica Ciomaga Hatnean, J. Ollivier, B. Winn, U. Filges, G. Balakrishnan, M. Kenzelmann, Nic Shannon and Tom Fennell, Nat. Phys. {\bf 14}, 711 (2018).

\bibitem{Gardner1999}
J. S. Gardner, S. R. Dunsiger, B. D. Gaulin, M. J. P. Gingras, J. E. Greedan, R. F. Kiefl, M. D. Lumsden, W. A. MacFarlane, N. P. Raju, J. E. Sonier, I. Swainson, and Z. Tun, Phys. Rev. Lett. {\bf 82}, 1012 (1999).

\bibitem{Jaubert2015}
L. D. C. Jaubert, Owen Benton, Jeffrey G. Rau, J. Oitmaa, R. R. P. Singh, Nic Shannon and Michel J. P. Gingras, Phys. Rev. Lett. {\bf 155}, 267208 (2015)

\bibitem{Yan2017}
Han Yan, Owen Benton, Ludovic Jaubert, and Nic Shannon, Phys. Rev. B {\bf 95}, 094422 (2017).

\bibitem{Martin2017}
N. Martin, P. Bonville, E. Lhotel, S. Guitteny, A. Wildes, C. Decorse, M. Ciomaga Hatnean, G. Balakrishnan, I. Mirebeau, and S. Petit
Phys. Rev. X {\bf 7}, 041028 (2017).

\bibitem{Wen2017}
J. J. Wen, S. M. Koohpayeh, K. A. Ross, B. A. Trump, T. M. McQueen, K. Kimura, S. Nakatsuji, Y. Qiu, D. M. Pajerowski, J. R. D. Copley and C. L. Broholm, Phys. Rev. Lett. {\bf 118}, 107206 (2017).

\bibitem{Hamid2007}
Hamid R. Molavian, Michel J. P. Gingras, and Benjamin Canals
Phys. Rev. Lett. {\bf 98}, 157204 (2007).

\bibitem{Princep2015}
A. J. Princep, H. C. Walker, D. T. Adroja, D. Prabhakaran, and A. T. Boothroyd, Phys. Rev. B {\bf 91}, 224430 (2015).

\bibitem{Huang2014} Y.-P. Huang, G. Chen, and M. Hermele, Phys. Rev. Lett. {\bf 112}, 167203 (2014).

\bibitem{Hatnean2015}
M. C. Hatnean, M. R. Lees, O. A. Petrenko, D. S. Keeble, G. Balakrishnan, M. J. Gutmann, V. V. Klekovkina, and B. Z. Malkin, Phys. Rev. B {\bf 91}, 174416 (2015).

\bibitem{Lhotel2015} E. Lhotel, S. Petit, S. Guitteny, O. Florea, M. Ciomaga Hatnean, C. Colin, E. Ressouche, M. R. Lees, and G. Balakrishnan, Phys. Rev. Lett. {\bf 115}, 197202 (2015).

\bibitem{Xu2015} J. Xu, V. K. Anand, A. K. Bera, M. Frontzek, D. L. Abernathy, N. Casati, K. Siemensmeyer, and B. Lake, Phys. Rev. B {\bf 92}, 224430 (2015).

\bibitem{Xu2016}
J. Xu, C. Balz, C. Baines, H. Luetkens, and B. Lake, Phys. Rev. B {\bf 94}, 064425 (2016).

\bibitem{Petit2016}
S. Petit, E. Lhotel, B. Canals, M. Ciomaga Hatnean, J. Ollivier, H. Mutka, E. Ressouche, A. R. Wildes, M. R. Lees, and G. Balakrishnan, Nat. Phys. {\bf 12}, 746 (2016).

\bibitem{Benton2016}
O. Benton, Phys. Rev. B {\bf 94}, 104430 (2016).

\bibitem{Opherden2017}
L. Opherden, J. Hornung, T. Herrmannsd\"orfer, J. Xu,
A. T. M. N. Islam, B. Lake, and J. Wosnitza, Phys. Rev. B {\bf 95}, 184418 (2017).

\bibitem{Xu2019}
J. Xu, Owen Benton, V. K. Anand, A. T. M. N. Islam, T. Guidi, G. Ehlers, E. Feng, Y. Su, A. Sakai, P. Gegenwart, and B. Lake
Phys. Rev. B {\bf 99}, 144420 (2019).

\bibitem{Lhotel2018}
E. Lhotel, S. Petit, M. Ciomaga Hatnean, J. Ollivier, H. Mutka, E. Ressouche, M. R. Lees, and G. Balakrishnan, Nat. Comm. {\bf 9}, 3786 (2018).

\bibitem{Xu2018}
J. Xu, A. T. M. N. Islam, I. N. Glavatskyy, M. Reehuis, J.-U. Hoffmann, and B. Lake, Phys. Rev. B {\bf 98}, 060408(R) (2018).

\bibitem{Ehlers2016}
G. Ehlers, A. A. Podlesnyak, and A. I. Kolesnikov, Rev Sci Instrum {\bf 87}, 093902 (2016).
\bibitem{Telling2016}
M. T. F. Telling and K. H. Andersen, Phys. Chem. Chem. Phys. {\bf 7}, 1255 (2004).
\bibitem{dave}
C.M. Brown, J.R.D. Copley, and R.M. Dimeo, J. Res. Natl. Inst. Stan. Technol. {\bf 114}, 341 (2009).
\bibitem{mantid}
O. Arnold et al., Nucl. Instrum. Methods Phys. Res. Sect. A {\bf 764}, 156–166 (2014).
\bibitem{horace}
R. A. Ewings, A. Buts, M. D. Le, J. van Duijn, I. Bustinduy, and T. G. Perring, Nuc. Ins. Methods Phys. Res. Sec. A: Accelerators, Spectrometers, Detectors and Associated Equipment, {\bf 834}, 132 (2016).

\bibitem{spinw} S. Toth and B. Lake, J. Phys.: Condens. Matter {\bf 27}, 166002 (2015).

\bibitem{SM}
Supplementary Materials.

\bibitem{Hao2014}
Z. Hao, A. G. R. Day and M. J. P. Gingras, Phys. Rev. B {\bf 90}, 214430 (2014).

\bibitem{Chang2012}
L.-J. Chang, S. Onoda, Y. Su, Y.-J. Kao, K.-D. Tsuei, Y. Yasui, K. Kakurai, and M. R. Lees, Nat. Commun. {\bf 3}, 992 (2012).

\bibitem{Powell2011}
S. Powell, Phys. Rev. B {\bf 84}, 094437 (2011).

\bibitem{Mauws2019}
C. Mauws, N. Hiebert, M. Rutherford, H. D. Zhou, Q. Huang, M. B. Stone, N. P. Butch, Y. Su, E. S. Choi, Z. Yamani, and C. R. Wiebe, arXiv:1906.10763 [cond-mat.str-el] (2019).

\bibitem{Mathieu2017}
Mathieu Taillefumier, Owen Benton, Han Yan, L. D. C. Jaubert, and Nic Shannon
Phys. Rev. X {\bf 7}, 041057 (2017).

\bibitem{Benton2018}
Owen Benton, L. D. C. Jaubert, Rajiv R. P. Singh, Jaan Oitmaa, and Nic Shannon
Phys. Rev. Lett. {\bf 121}, 067201 (2018).

\bibitem{Anand2015}
V. K. Anand, A. K. Bera, J. Xu, T. Herrmannsd\"{o}rfer, C. Ritter, and B. Lake, Phys. Rev. B {\bf 92}, 184418 (2015).

\bibitem{Bertin2015}
A. Bertin, P. Dalmas de R\'eotier, B. F\aa{}k, C. Marin, A. Yaouanc, A. Forget, D. Sheptyakov, B. Frick, C. Ritter, A. Amato, C. Baines, and P. J. C. King, Phys. Rev. B {\bf 92}, 144423 (2015).

\bibitem{Hallas2015}
A. M. Hallas, A. M. Arevalo-Lopez, A. Z. Sharma, T. Munsie, J. P. Attfield, C. R. Wiebe, and G. M. Luke, Phys. Rev. B {\bf 91}, 104417  (2015).

\bibitem{Mauws2018}
C. Mauws, A. M. Hallas, G. Sala, A. A. Aczel, P. M. Sarte, J. Gaudet, D. Ziat, J. A. Quilliam, J. A. Lussier, M. Bieringer, H. D. Zhou, A. Wildes, M. B. Stone, D. Abernathy, G. M. Luke, B. D. Gaulin, and C. R. Wiebe, Phys. Rev. B {\bf 98}, 100401(R) (2018).

\bibitem{Viviane2019}
Viviane Pe\c canha-Antonio, Erxi Feng, Xiao Sun, Devashibhai Adroja, Helen C. Walker, Alexandra S. Gibbs, Fabio Orlandi, Yixi Su, and Thomas Brückel, Phys. Rev. B {\bf 99}, 134415 (2019).

\end{thebibliography}
\end{document}